\documentclass[a4paper]{article}

\pdfoutput=1

\usepackage{epsfig,amsfonts,amsthm}
\usepackage{amsmath,amssymb,color}
\usepackage{epstopdf}
\usepackage{jheppub}

\newcommand{\be}{\begin{equation}}
\newcommand{\ee}{\end{equation}}
\newcommand{\ba}{\begin{align}}
\newcommand{\ea}{\end{align}}
\newcommand{\bea}{\begin{eqnarray}}
\newcommand{\eea}{\end{eqnarray}}
\newcommand{\no}{\nonumber\\}

\newcommand{\Z}{\mathbb{Z}}

\def\lsim{\mathrel{\rlap{\lower4pt\hbox{\hskip1pt$\sim$}}
    \raise1pt\hbox{$<$}}}         
\def\gsim{\mathrel{\rlap{\lower4pt\hbox{\hskip1pt$\sim$}}
    \raise1pt\hbox{$>$}}}         

\begin{document}

\title{Electroweak vacuum lifetime in two Higgs doublet models}

\author[a,b]{V. Branchina}
\author[a,b,c]{F. Contino}
\author[d,e]{P.M.~Ferreira}

\affiliation[a]{Department of Physics and Astronomy, University of Catania,
Via Santa Sofia 64, 95123 Catania, Italy}
\affiliation[b]{INFN, Sezione di Catania, Via Santa Sofia 64, 95123 Catania, Italy}
\affiliation[c]{Scuola Superiore di Catania, Via Valdisavoia 9, 95123 Catania, Italy}
\affiliation[d]{Instituto Superior de Engenharia de Lisboa- ISEL,
	1959-007 Lisboa, Portugal}
\affiliation[e]{Centro de F\'{\i}sica Te\'{o}rica e Computacional,
    Faculdade de Ci\^{e}ncias,
    Universidade de Lisboa,
    Av.\ Prof.\ Gama Pinto 2,
    1649-003 Lisboa, Portugal}

\emailAdd{vincenzo.branchina@ct.infn.it}
\emailAdd{filippo.contino@ct.infn.it}
\emailAdd{pmmferreira@fc.ul.pt}

\date{\today}

\abstract{
We study the stability of neutral electroweak vacua in two Higgs doublet models, and calculate
the lifetime of these states when the parameters are such that they are false vacua.
As the two Higgs doublet model is invariant under a sign change of both doublets, degenerate true
vacua exist. It is shown that this degeneracy, despite the fact that each of these minima locally describes
 the same physics, can immensely affect their lifetime.
We apply these calculations to the parameter space of the models
which is allowed by recent LHC searches, and infer combinations of parameters which should be excluded on grounds of
a tunneling time inferior to the age of the universe.
}

\maketitle

\section{Introduction}

In 2012 the LHC discovered the last missing piece of the Standard Model (SM) of particle physics~\cite{Aad:2012tfa,Chatrchyan:2012ufa}, the Higgs boson.
We are now measuring with increasing precision the properties of this new particle, thus
probing the hitherto unknown scalar sector of the theory. The trend in the ATLAS and CMS
results is clear: the 125 GeV scalar discovered behaves very much like the SM Higgs boson
was expected to behave (see, for instance, \cite{Khachatryan:2016vau}). However, the current
experimental results for the scalar sector still leave a lot of room for Beyond Standard Model
(BSM) physics to occur, and BSM is required to explain a great number of phenomena which the SM
cannot account for, of which the origin of dark matter, dark energy and the matter-antimatter asymmetry
are but three amongst many other issues to be resolved.

The two-Higgs doublet model (2HDM)~\cite{Lee:1973iz,Branco:2011iw} is arguably the simplest SM extension,
in which the particle content of the SM is complemented by a second Higgs doublet. The model boasts a
rich phenomenology, with a larger scalar sector, including two CP-even scalars, a pseudoscalar and a charged
scalar, and may have spontaneous CP breaking for certain choices of its parameters, thus offering an
additional source of CP violation which might help in explaining  why the universe contains so
much more matter than antimatter. In one version of the 2HDM -- the so called {\em inert model}
\cite{Ma:1978,Barbieri:2006,Cao:2007,LopezHonorez:2006} -- dark matter candidates arise naturally, their
lack of interactions with ``normal" matter being ensured by a discrete symmetry unbroken by the vacuum.
The 2HDM does an excellent job at fitting the existing experimental data. The properties of
the discovered scalar, very similar to those expected within the
SM, can easily be reproduced in the 2HDM, as seen soon after discovery~\cite{Ferreira:2011aa,Ferreira:2012my}
and is today easily verified for much of the parameter space using, for instance, the
{\tt HiggsSignals} code \cite{Bechtle:2013xfa}.
Of course, the 2HDM predicts the existence of other scalars, as yet not discovered, so the model must also
be in agreement with current
experimental searches for BSM particles. Even after demanding that the
125 GeV scalar be SM-like, there remains a large 2HDM
parameter space available to comply with those experimental results (as proven by application of the
{\tt HiggsBounds} code \cite{Bechtle:2008jh,Bechtle:2011sb,Bechtle:2013wla}).
Also, the model can easily fit constraints on the charged Higgs mass arising from other
observables, such as B-meson physics~\cite{Deschamps:2009rh,Mahmoudi:2009zx,Hermann:2012fc,
Misiak:2015xwa,Misiak:2017bgg,Arbey:2017gmh,Haber:1999zh}.

Another interesting property of the 2HDM is that already at the classical level it has a richer vacuum
structure -- whereas in the classical SM potential there can only
be one type of minimum, the 2HDM has the possibility of {\em three} physically different kinds of minima:
an electroweak-breaking but CP-and-charge preserving (we call it ``normal" minimum), analogous to the SM;
a minimum which spontaneously breaks
both the electroweak and CP symmetries; and a minimum where the vacuum expectation value (vev) of the scalar
doublets carries electric charge, and electric charge conservation no longer holds. However, the scalar potential of
the model is such that, at least at tree level, minima of a different nature cannot simultaneously coexist
\cite{Ferreira:2004yd,Barroso:2005sm,Ivanov:2006yq,Ivanov:2007de}. In other words, if for instance a normal minimum exists,
 which is nothing more than an electroweak (EW)
minimum which preserves charge and CP, any possible charge breaking (CB) or CP breaking stationary
points of the potential are necessarily saddle points, and in addition lie above the EW minimum. Similar conclusions
are valid for CB or CP minima - if they exist, all other types of possible stationary points will
be saddle points lying above them. The stability of a 2HDM vacuum against tunneling to another vacuum of a different
nature is therefore ensured by the theory itself, at least at tree level.

There is however another 2HDM property concerning normal vacua: for certain regions of the parameter space,
there may exist {\em two} non-degenerate vacua of this type~\cite{Barroso:2007rr,Ivanov:2006yq,Ivanov:2007de},
both of them CP and charge preserving, but having vevs which break the electroweak symmetry. However, the value
of those vevs is different for each minimum, which means that, since all elementary particles gain their mass
from interactions with scalar particles, the mass spectrum at each of the two minima is quite different. In other
words, the vevs of the doublets, $v_i$, are such that in ``our" minimum they satisfy $v_1^2 + v_2^2 = 246$ GeV$^2$
-- thus in ``our" minimum all elementary particles have their well-known masses -- whereas a different mass spectrum holds
for the second minimum. If the EW vacuum which the universe currently occupies is not the absolute
minimum of the potential, it will sooner or later  tunnel to a deeper minimum that breaks the same symmetries.
The (surprisingly simple) conditions under which a second 2HDM minimum may exist, and the condition
which discriminates whether ``our" minimum is the global one were established in
refs.~\cite{Ivanov:2006yq,Ivanov:2007de,Barroso:2012mj,Barroso:2013awa}. The deeper vacuum, different
from the ``standard" EW breaking one, was dubbed {\em panic vacuum}
in~\cite{Barroso:2012mj,Barroso:2013awa}: a transition from the EW minimum to the deeper one would be
disastrous. In fact, such a transition  would release a colossal amount of energy; and since the fields
in the two minima have different vevs, all elementary particles would change their masses upon transition
to the deeper vacuum. Both situations are nothing short of catastrophic.

The mere existence of a deeper, ``panic" vacuum, however, is not sufficient to exclude the parameters of
the potential which yield such a bizarre possibility. In fact, if the tunneling time $\tau$ from the false to
the true vacuum
is larger than the age of the universe, the existence of the deeper vacuum would have no
impact whatsoever in the phenomenology observed while the universe lies in the upper minimum. Thus
the computation of $\tau$ becomes a fundamental tool to distinguish between
those regions of the parameter space which yield dangerous panic vacua, and those for which the deeper
vacua exist but are practically harmless.

It is now worth it to stress an important difference between the 2HDM and the SM. In the latter case the
EW vacuum shows its instability (metastability) only once radiative corrections are
taken into account, and this is mainly due to the negative contribution to the potential coming from the top
quark. Actually the Higgs effective potential $V(\phi)$ turns over for values of $\phi >\, v = 246$ GeV, and
for the present experimental values of $m_h$ and $m_t$, namely $m_h \sim 125.09$ GeV and $m_t \sim 173.34$ GeV
\,\cite{Aad:2015zhl,ATLAS:2014wva}, it develops a second minimum much deeper than the EW one and at a much larger
value of the field. Under these conditions, then, the EW minimum is a metastable state, a so called ``false vacuum''
($\phi_{\rm fv} \equiv v$), while the deeper minimum is the true vacuum (that occurs at
$\phi_{\rm tv}\, \gg \, v$. Unlike the SM case, however, the coexistence of two minima in
the 2HDM potential already occurs at tree-level, and the analysis of the stability of the false vacuum (the EW minimum
in our case) can be already undertaken at this level (i.e. prior to the study of the loop corrected potential),
as was the case for the pioneering work of Coleman and collaborators~\cite{Coleman:1977py,Callan:1977pt},
where a (classical) potential
with two minima of different depths was considered.

Physically the false vacuum decay is triggered by quantum fluctuations, that induce a finite probability for a
bubble of true vacuum to materialize in a false vacuum sea. Both in flat and curved spacetime backgrounds,
Coleman and collaborators considered a scalar theory where the potential $V(\phi)$ has a relative and an
absolute minimum, at $\phi_{\rm fv}$ and $\phi_{\rm tv}$ respectively, such that the  energy density
difference $V(\phi_{\rm fv}) - V(\phi_{\rm tv})$ is much smaller than the height of the
``potential barrier'' $V(\phi_{\rm top}) - V(\phi_{\rm fv})$, where $V(\phi_{\rm top})$ is the maximum
of the potential between the two minima. Under these conditions  the true vacuum bubble is separated from
the false vacuum sea by a ``thin wall'', and this allows us to treat the problem analytically, within the so
called ``thin wall'' approximation.

The conditions under which this approximation can be applied are however not fulfilled in the SM case, at
least for the central values of the Higgs and top masses reported above, but this is not a problem because
the stability analysis can be performed numerically. If the EW vacuum is metastable, the fate of our
universe is to decay sooner or later into the true vacuum, and it is then of the greatest importance to
estimate its lifetime $\tau$. When the SM alone is considered, the masses of the Higgs boson and of the
top quark are taken as reported above, and it is also assumed that interactions at higher energy scales
have no impact on the stability condition of the EW vacuum, $\tau$ turns out to be much larger than the
age of the universe $T_U$\,\cite{EliasMiro:2011aa,Degrassi:2012ry}. Actually
$\tau \sim 10^{640} \, T_U$ ($T_U \sim 13,7 \cdot 10^9$ years), and accordingly we can say that our
universe is practically stable.

It was later realized, however, that the stability condition of the EW vacuum is very sensitive to
unknown New Physics (even if that physics occurs at high energy scales), and the decay rate of the
EW vacuum can be strongly modified by its
presence\,\cite{Branchina:2013jra,Branchina:2014usa,Branchina:2014rva,Branchina:2016bws,Branchina:2015nda,Bentivegna:2017qry}.
This leads to the conclusion
that  models of BSM physics should not only satisfy all
current experimental constraints stemming from precision measurements,
but should also be tested against a careful stability analysis, as only models for which the EW vacuum
lifetime turns out to be larger than the age of the universe are physically acceptable.

In this paper, we will undertake a thorough analysis of the tunneling between neutral vacua in the 2HDM,
by calculating the tunneling time from false to true vacua.
To this end, we have to look for the so-called bounce solutions to the Euclidean Euler-Lagrange equations that
have O(4) symmetry and satisfy specific boundary conditions~\cite{Coleman:1977py}. In principle these bounces
are composed by eight fields, since the two doublets have eight real components, unlike the SM case where only
one field is present. Using gauge invariance arguments, however, we will show that the problem is reduced to a
five-field calculation.
Analysing the bounce equations, from which tunneling times are computed, we will show that
for CP-conserving potentials the problem is further reduced to a two-field calculation -- but we will
also show that for potentials with explicit CP violation this is no longer the case, and in general the
bounce solutions will involve three fields. Further consideration of the full minimum landscape of the 2HDM
leads to the conclusion that the lifetime of false vacua may be enormously affected by the existence of degenerate true vacua,
which is caused by the invariance of the potential under a sign swap of both doublets.
Using a dedicated and very efficient code to compute tunneling
times in theories with many fields~\cite{Masoumi:2016wot}, we will probe large regions of 2HDM
parameter space -- complying with all theoretical and experimental constraints that the model is expected
to obey in this LHC era -- and verify under which conditions dangerous deeper minima develop. We will
show that specific benchmarks of the model -- for which we specify 6 out of 8 of the scalar sector
parameters -- may be completely safe, boasting either a single minimum
or tunneling times to deeper minima far larger than the age of the universe. Nonetheless, other perfectly banal
benchmarks may have over 11\% of the remaining 2-parameter space excluded on account of having
far too short tunneling times to a deeper vacuum.

\section{The Two-Higgs Doublet Model potential}
\label{sec:2hdm}

The 2HDM is perhaps the simplest extension of the SM  -- the particle content of the 2HDM is enlarged by
a second $SU(2)_W\times U(1)_Y$ doublet, but the gauge and fermion content of the
model is the same as the SM's. The model was proposed by T.D. Lee in 1973~\cite{Lee:1973iz}
as a means to obtain CP violation from spontaneous symmetry breaking. For a review, see
\cite{Branco:2011iw}. The model therefore contains two hypercharge $1$ doublets, $\Phi_1$ and $\Phi_2$,
in terms of which the most general renormalizable 2HDM scalar potential is written as
\bea
V &=&
m_{11}^2 |\Phi_1|^2
+ m_{22}^2 |\Phi_2|^2
- \left( m_{12}^2 \Phi_1^\dagger \Phi_2 + h.c. \right)
\no & &
+ \frac{1}{2} \lambda_1 |\Phi_1|^4
+ \frac{1}{2} \lambda_2 |\Phi_2|^4
+ \lambda_3 |\Phi_1|^2 |\Phi_2|^2
+ \lambda_4 |\Phi_1^\dagger\Phi_2|^2
\no & &
+ \left[
\frac{1}{2} \lambda_5 \left( \Phi_1^\dagger\Phi_2 \right)^2
+ \lambda_6 |\Phi_1|^2
\left( \Phi_1^\dagger\Phi_2 \right)
+ \lambda_7 |\Phi_2|^2
\left( \Phi_1^\dagger\Phi_2 \right)
+ h.c. \right],
\label{eq:pot}
\eea
where the coefficients $m_{12}^2$, $\lambda_{5,6,7}$ can be complex. The doublets $\Phi_1$ and $\Phi_2$ are
not physical fields -- the mass eigenstates which arise from them are physical, but the doublets themselves
are not. This means that any linear combination of the doublets which preserves the form of the model's
kinetic terms provides an equally valid physical description of physics -- this corresponds to an invariance
of the model under fields redefinitions, so called {\em basis changes} of the form
$\Phi^\prime_i = U_{ij} \Phi_j$, where $U$ is a $2\times 2$ unitary matrix. Though the potential of
eq.~\eqref{eq:pot} seemingly has 14 independent real parameters, the freedom to redefine
the doublets  means that in fact one can eliminate three of those
parameters, and thus the most general 2HDM scalar potential has 11 independent real
parameters~\cite{Gunion:2002zf}.

When considering the whole theory we must include scalar-fermion interactions -- the Yukawa
sector. And there we run into a problem -- if we build the most general
lagrangian with two Higgs doublets, the Yukawa sector will include tree-level
flavour changing neutral currents (FCNC) mediated by neutral scalars. This happens because the
most general Yukawa terms of the 2HDM include interactions of both doublets with all fermions.
However, these FCNC are very tightly constrained by experimental data and they should be avoided. This
may be achieved, of course, by simply fine tuning the Yukawa couplings -- there is sufficient freedom
in the Yukawa coupling matrices to achieve this. One other possibility is to assume an ``alignment" {\em ansatz}
relating Yukawa matrices~\cite{Pich:2009sp,Jung:2010ik,Jung:2010ab,Ferreira:2010xe,Gori:2017qwg}.
The most studied model, however, eliminates tree-level
scalar-mediated FCNC by imposing a ${\Z}_2$ discrete symmetry upon the model -- this method, unlike the previous
ones, is entirely stable under renormalization. The discrete symmetry usually considered demands that the
lagrangian be invariant under a transformation on the doublets of the form $\Phi_1 \rightarrow \Phi_1$ and
$\Phi_2 \rightarrow -\Phi_2$ \cite{Glashow:1976nt,Paschos:1976ay}. As a
consequence, the parameters $m_{12}^2$, $\lambda_6$ and $\lambda_7$ vanish from
the potential -- though $m_{12}^2$ is reintroduced as a (real) soft-breaking term, to enlarge
the allowed parameter space and, among other things, allow the theory to have a {\em decoupling limit}
\cite{Gunion:2002zf} where the masses of all scalars other than the SM-like one can be made very large.
The final potential with which we will be working is thus
\bea
V &=&
m_{11}^2 |\Phi_1|^2
+ m_{22}^2 |\Phi_2|^2
- m_{12}^2 \left(\Phi_1^\dagger \Phi_2 + h.c. \right)
\no & &
+ \frac{1}{2} \lambda_1 |\Phi_1|^4
+ \frac{1}{2} \lambda_2 |\Phi_2|^4
+ \lambda_3 |\Phi_1|^2 |\Phi_2|^2
+ \lambda_4 |\Phi_1^\dagger\Phi_2|^2
+
\frac{1}{2} \lambda_5 \left[ \left( \Phi_1^\dagger\Phi_2 \right)^2
+ h.c. \right],
\label{eq:pot2}
\eea
where now all parameters are real (we have further imposed CP conservation on the
potential, which makes all possible complex phases vanish).

The 2HDM, of course, is not only a theory of the scalar sector, it includes also gauge bosons and
three generations of fermions, as does the SM. The most general Yukawa sector of the model, as mentioned
above, will generate tree-level FCNC which are strongly disfavoured by experimental results. These
are eliminated imposing, on the full lagrangian, the discrete symmetry $\Phi_1 \rightarrow \Phi_1$ and
$\Phi_2 \rightarrow -\Phi_2$ and we have already explained the impact
of this symmetry on the scalar sector; on the Yukawa sector, it forces only one of the doublets to couple
(and thus give mass) to each generation of like-charged fermions. Depending on how the fermionic fields
(both the left doublets and right singlets) transform under the  ${\Z}_2$ symmetry, there are then
several possible types of 2HDM, with different phenomenologies and classified according to their
scalar-fermion interactions. Usually, one considers four different types~\footnote{The number of
possible models would increase if one were to consider also the possible interaction terms between
the scalar doublets and neutrinos, which we will not do in the current work.}:
\begin{itemize}
\item Model Type I, where all fermions couple to a single Higgs doublet, chosen as $\Phi_2$ per
convention.
\item Model Type II, where all right-handed up-type quarks couple to $\Phi_2$, but right-handed
down-type quarks and charged leptons couple to $\Phi_1$. This type of couplings is analogous to what
happens in SUSY models.
\item The Lepton-specific model, in which all quarks couple to $\Phi_2$, but right-handed charged
leptons couple to $\Phi_1$.
\item The Flipped model, in which right handed up quarks and charged leptons couple to
$\Phi_2$, but right-handed down quarks couple to $\Phi_1$.
\end{itemize}
Thus for each model each same-charge type of fermions may gain their masses from different Higgs
doublets. The fact that only one Higgs doublet couples to fermions of the same electric charge
eliminates tree-level FCNC, as the couplings between the physical scalar particles and the fermions
will be described by diagonal matrices~\cite{Branco:2011iw}. As already mentioned, each of these
models has different phenomenologies, a subject we will address in section~\ref{sec:exp}.

\subsection{Theoretical constraints on quartic couplings}

Notice that the quartic couplings of
\eqref{eq:pot2} are not completely unconstrained -- in order
to ensure that the potential is bounded from below (BFB), meaning, no directions in field space
along which the potential can tend to minus-infinity, the couplings need to obey
\cite{Deshpande:1977rw}
\bea
\lambda_1 > 0 & , &  \lambda_2 > 0 \; ,\nonumber \\
\lambda_3 > -\sqrt{\lambda_1 \lambda_2} & , &
\lambda_3 + \lambda_4 - |\lambda_5| > -\sqrt{\lambda_1 \lambda_2} \;.
\label{eq:bfb}
\eea
It has been proven that these (tree-level) conditions are both necessary and
sufficient \cite{Ivanov:2006yq,Ivanov:2007de}. It is possible to go beyond tree-level
in these BFB constraints -- this is usually accomplished by studying the renormalization group
evolution of the quartic couplings of the potential and imposing that the conditions shown in eqs.~\eqref{eq:bfb}
be valid at all scales. Prior to the discovery of the Higgs boson, this procedure was used to constrain the
2HDM parameter space (see, for instance, ~\cite{Kreyerhoff:1989fa,Nie:1998yn,Kanemura:1999xf,Ferreira:2009jb}).
Post-Higgs discovery, this method has shown that the metastability claimed for the SM effective
potential~\cite{Bezrukov:2012sa, Degrassi:2012ry, Buttazzo:2013uya}, which seemingly may develop a deeper
minimum if the theory is considered valid all the way up to the Planck scale, may be cured in the 2HDM due to its
larger scalar content~\cite{Chakrabarty:2014aya,Das:2015mwa,Chowdhury:2015yja,Ferreira:2015rha,
Chakrabarty:2016smc,Cacchio:2016qyh,Chakrabarty:2017qkh,Gori:2017qwg,Basler:2017nzu}.
In the current work we will confine ourselves to tree-level conditions~\footnote{In any case, since we will
only consider regions of the 2HDM well within the so-called ``alignment limit", we expect that the tree-level
conditions will be more than sufficient for the model to be valid up to very high scales~\cite{Basler:2017nzu}.}.
Another set of constraints upon the potential's parameters arises from requiring that
the theory be unitary -- this translates into further constraints upon the quartic couplings
of the potential, which may be reduced to~\cite{Kanemura:1993hm, Akeroyd:2000wc,Horejsi:2005da}
\bea
|\lambda_3 - \lambda_4| &<& 8 \pi \label{eq:ev1} \nonumber \\
|\lambda_3 + 2 \lambda_4 \pm 3 \lambda_5| &<& 8 \pi \nonumber \\
\left| \frac{1}{2} \left( \lambda_1 + \lambda_2 + \sqrt{(\lambda_1 -
     \lambda_2)^2 + 4 \lambda_4^2}\right) \right| &<& 8\pi \nonumber \\
\left| \frac{1}{2} \left( \lambda_1 + \lambda_2 + \sqrt{(\lambda_1 -
     \lambda_2)^2 + 4 \lambda_5^2}\right) \right| &<& 8\pi. \label{eq:ev5}
\eea
Again, we will consider tree-level unitarity constraints,
though one-loop contributions have been considered, and shown to curtail the available
2HDM parameter space~\cite{Grinstein:2015rtl,Cacchio:2016qyh}.

\subsection{The electroweak-breaking minimum}

The potential described by eq.~\eqref{eq:pot2} can yield, depending of the values of the parameters,
different types of minima. The scalar fields can acquire vacuum expectation values (vevs) and break
the symmetries of the model in different ways. We call ``normal vacuum" the case where both doublets
acquire real and neutral vevs,
\be
\langle\Phi_1 \rangle_N = {\displaystyle\frac{1}{\sqrt{2}}}\begin{pmatrix} 0 \\ v_1 \end{pmatrix} \; , \;
\langle\Phi_2 \rangle_N = {\displaystyle\frac{1}{\sqrt{2}}}\begin{pmatrix} 0 \\ v_2 \end{pmatrix}.
\label{eq:vevn}
\ee
These normal minima are similar to the SM vacuum -- they break the same gauge symmetries and preserve CP,
and constitute the focus of the work of this paper (we will briefly discuss other types of possible
2HDM minima in section~\ref{sec:min}). Let us now define the (real) components of the doublets $\Phi_1$ and $\Phi_2$
as
\be
\Phi_1 \,=\,\frac{1}{\sqrt{2}}\,\left(
\begin{array}{c} \varphi_{c1} + \mbox{i} \,\varphi_{c2} \\ \varphi_{r1} + \mbox{i} \,\varphi_{i1} \end{array}
\right)\;\;\; , \;\;\;
\Phi_2 \,=\,\frac{1}{\sqrt{2}}\,\left(
\begin{array}{c} \varphi_{c3} + \mbox{i} \,\varphi_{c4} \\ \varphi_{r2} + \mbox{i} \,\varphi_{i2} \end{array}
\right)\;,
\label{eq:Phi}
\ee
where the upper components correspond to charged (+1) fields and the lower components, to neutral ones.
When the potential develops a normal minimum, the real neutral components, $\varphi_{r1}$ and $\varphi_{r2}$,
give rise to two mass eigenstates which correspond to CP-even scalars, dubbed $h$ and $H$. On the other hand,
the imaginary components, $\varphi_{i1}$ and $\varphi_{i2}$, originate a pseudoscalar particle, $A$, and the
neutral Goldstone boson $G^0$ which provides the $Z$ boson with its mass. Finally, the upper, charged components
$\varphi_{ci}$ yield a charged Higgs scalar, $H^\pm$ and the charged Goldstone boson $G^\pm$ which gives mass to the
$W$ gauge bosons. For such normal minima it is also customary to define two angles: the ratio of the vevs $v_1$ and
$v_2$ defines the angle $\beta$, such that
\be
\tan\beta\,=\,\frac{v_2}{v_1}\,.
\ee
$\beta$ is the angle which diagonalizes both the charged and pseudoscalar squared scalar mass matrices, and
can be considered to only take values between $0$ and $\pi/2$ without loss of generality~\footnote{This choice
is valid for one specific vacuum, other vacua may have vevs of different signs.}. On the other hand,
the CP-even squared scalar mass matrix is diagonalized by a different angle, $\alpha$, defined such that the
two physical eigenstates, $h$ and $H$, are related to the neutral real components of the doublets as
\bea
h &=& \sin\alpha \,\varphi_{r1}\,-\,\cos\alpha \,\varphi_{r2}\nonumber \\
H &=& - \cos\alpha \,\varphi_{r1}\,-\,\sin\alpha \,\varphi_{r2}\,.
\eea
Again without loss of generality, this angle can be chosen such that $-\pi/2 \leq \alpha \leq \pi/2$.
The minimization conditions relate the vevs of eq.~\eqref{eq:vevn} to the parameters of the potential,
such that
\bea
m_{11}^2 v_1\,-\,m_{12}^2 v_2\,+\,\frac{\lambda_1}{2}v_1^3\,+\,\frac{\lambda_{345}}{2}v_2^2 v_1 &=& 0
\nonumber \\
m_{22}^2 v_2\,-\,m_{12}^2 v_1\,+\,\frac{\lambda_2}{2}v_2^3\,+\,\frac{\lambda_{345}}{2}v_1^2 v_2 &=& 0\;,
\label{eq:min}
\eea
where we have defined
\be
\lambda_{345} \equiv \lambda_3 + \lambda_4 + \lambda_5 \;.
\ee
Notice that, since the potential is invariant under a sign change for both doublets, if eqs.~\eqref{eq:min}
admit a solution $\{v_1\,,\,v_2\}$ obviously $\{-v_1\,,\,-v_2\}$ will also be a solution.
Also obviously, this second solution will be physically indistinguishable from the first one. This
seemingly trivial point will be extremely important later on, and we will show in section~\ref{sec:mindeg}
that it can have a stunning impact on the tunneling rates between vacua.

Instead of the potential's couplings, we can choose to describe the model in terms of the four physical
masses, $m_h = 125$ GeV, $m_H$, $m_A$ and $m_{H^\pm}$, the angles $\beta$ and $\alpha$, the vev $v = 246$
GeV and a further parameter, for instance the soft breaking term $m^2_{12}$ -- a total of eight parameters,
just as the potential of eq.~\eqref{eq:pot2}. The quartic couplings of the model can then be expressed as
\bea
\lambda_1
&=&
\frac{1}{v^2 c_\beta^2}\left(c_\alpha^2 m_H^2 + s_\alpha^2 m_h^2
- m^2_{12}\frac{s_\beta}{c_\beta}\right) , \no
\lambda_2
&=&
\frac{1}{v^2 s_\beta^2}\left( s_\alpha^2 m_H^2 + c_\alpha^2 m_h^2
- m^2_{12}\frac{c_\beta}{s_\beta}\right), \no
\lambda_3
&=& \frac{1}{v^2} \left[
2 m_{H^\pm}^2 +
\frac{s_{2 \alpha} (m_H^2 - m_h^2)}{s_{2 \beta}} - \frac{m^2_{12}}{s_\beta c_\beta}
\right],
\no
\lambda_4
&=& \frac{1}{v^2} \left(
m_A^2 - 2 m_{H^\pm}^2 + \frac{m^2_{12}}{s_\beta c_\beta}\right), \no
\lambda_5
&=& \frac{1}{v^2} \left( \frac{m^2_{12}}{s_\beta c_\beta} - m_A^2 \right)\;,
\label{eq:coup}
\eea
where for simplification we defined $s_\theta = \sin\theta$ and $c_\theta = \cos\theta$.

\subsection{Experimental constraints on the 2HDM}
\label{sec:exp}

The larger scalar content of the 2HDM, compared with the SM, leads to measurable
impacts on several experimental observables. So far no scalars other than the 125 GeV one
have been discovered -- and therefore BSM searches at the LHC and elsewhere impose bounds
on the masses and couplings of the extra scalars of the 2HDM. Further, even before the discovery
of the Higgs boson, electroweak precision studies from LEP and other accelerators were used
to curtail the values of BSM models, including the 2HDM. A charged scalar such as the
one predicted by the 2HDM has considerable contributions to several B-meson observables, and
data from B-physics measurements constitute some of the model's most stringent constraints.
In the current work we incorporated a wealth of experimental constraints in the parameter
scans used in section~\ref{sec:res}.

In general, BSM physics may have substantial contributions to Electroweak Precision Constraints
(EWPC), namely the oblique $S$, $T$ and $U$
parameters~\cite{Peskin:1990zt,Peskin:1991sw,Maksymyk:1993zm}. These constraints
may, for instance, force the charged Higgs mass and the pseudoscalar one to be very close in value.
We computed these oblique parameters and used the most recent fit~\cite{Baak:2014ora} to constrain
the 2HDM parameter space. Direct searches from LEP, using the channel
$e^+ e^- \to H^+ H^-$~\cite{Abbiendi:2013hk}, impose a lower bound on the charged Higgs mass of
roughly 80 GeV, which we also implemented~\cite{Arbey:2017gmh}. And, as described above, the 2HDM contributions to
B-physics observables, such as the values of the $b\rightarrow s\gamma$ decay
rate~\cite{Deschamps:2009rh,Mahmoudi:2009zx,Hermann:2012fc,Misiak:2015xwa,Misiak:2017bgg} and
the $Z\rightarrow b\bar{b}$ width~\cite{Haber:1999zh,Deschamps:2009rh},
impose considerable constraints, usually expressed as exclusions on the
$m_{H^\pm}$--$\tan\beta$ plane. Roughly speaking, these constraints translate as requiring that
$\tan\beta$ be above $1$ for most of the parameter space in all model types, and an almost $\tan\beta$-independent
lower bound on the charged Higgs mass for model type II (and Flipped), of roughly
$\sim 580$ GeV~\cite{Misiak:2017bgg}. Other flavour constraints, such as those arising from
$B\rightarrow \tau \nu$, $\Delta M_{B_{s,d}}$, etc.~\cite{Arbey:2017gmh}, were also taken into account.

The Higgs boson discovery at the LHC has been followed by many measurements of this particle's properties,
which have been seen to be very much in agreement with what one could expect for a SM-like scalar. The
experimental results are thus pushing the 2HDM into the so-called ``alignment limit", wherein the 125 GeV
state is almost ``aligned" with one of the doublets (this in practice corresponds to values of $\sin(\beta
-\alpha)$ very close to 1), and the remaining scalars sufficient massive, or with sufficiently weak interactions,
to have eluded detection thus far. In practical terms, the LHC constraints are obtained from the $\mu$ ratios between the
observed number of events in some Higgs-mediated channel, and the SM expected value for the same quantity. For the
2HDM, then, the quantities to compare with experimental results such as those from~\cite{Khachatryan:2016vau}
are
\be
\mu_X\,=\,\frac{\sigma^{2HDM}(pp\rightarrow h)}{\sigma^{SM}(pp\rightarrow h)}\,
\frac{BR^{2HDM}(h\rightarrow X)}{BR^{SM}(h\rightarrow X)}\, ,
\label{eq:muX}
\ee
where $\sigma$ stands for the production cross section of $h$ in proton-proton collisions at the LHC
and $BR$ for the decay branching ratios of $h$ to some final state $X$, such as $ZZ$, $WW$, $\gamma\gamma$, $b\bar{b}$,
$\dots$. The fact that $h$ is behaving in a SM-like manner means that the measured values for these $\mu_X$ are
close to one, but the current experimental uncertainties still allow values with deviations larger than
30\% from unity. In our calculations we will consider mostly scalars produced via the main channel of
gluon-gluon fusion, the cross sections of such processes being obtained by {\tt SusHiv1.6.0}
\cite{Harlander:2012pb,Harlander:2016hcx}, at NNLO QCD. Other production channels (such as VBF, $b\bar{b}h$
or $t\bar{t}h$) were also computed, but since they are subdominant, for the purposes of the current paper
we chose not to use them. As for the branching ratios, all decay widths were computed at leading order, with
the necessary NLO QCD corrections to the $b\bar{b}$ width taken into account. In fact, requiring that $\mu_{ZZ}$, $\mu_{\gamma\gamma}$, $\mu_{b\bar{b}}$ and
$\mu_{\tau\bar{\tau}}$ be within 30\% of their SM value ({\em i.e.}, all $\mu$'s having values in the
interval 0.7 to 1.3) is enough to have a rough compliance with the $2\times 1\sigma$ experimental
precision from~\cite{Khachatryan:2016vau}.

Finally, there is a wealth of results on searches for the extra scalars predicted in the 2HDM (see
ref.~\cite{Han:2017pfo} and references therein, for a review of the status of the diverse search
channels), with measurements imposing exclusion regions in the parameter space of the model. By and large,
requiring that the 125 GeV state $h$ be very SM-like is sufficient to comply with most exclusion bounds for
other scalar searches, even though there are exceptions~\cite{Ferreira:2017bnx}, like pseudoscalar production
and decay to $Zh$ in the wrong sign limit in the
2HDM~\cite{Ferreira:2014naa,Ferreira:2014dya,Dumont:2014wha,Fontes:2014tga,Bernon:2014nxa,Biswas:2015zgk,
Modak:2016cdm}. For the purposes of the current work, in which we wish to show the possible importance of the
tunneling time calculations in 2HDM parameter space, we have verified that in regions of parameters analysed
the 30\% bound on the several $\mu_X$ was sufficient to comply with extra scalar search results.

\section{Coexisting minima in the 2HDM}
\label{sec:min}

Since the 2HDM has a scalar potential much more elaborate than the SM one, it possesses therefore a richer
vacuum structure. In fact, in the 2HDM {\em three} classes of vacua may occur, depending on the parameters of the model.
The first corresponds to  normal
vacua, wherein the doublets have vevs such as those described by eq.~\eqref{eq:vevn}. This kind of vacuum
therefore breaks $SU(2)_L\times U(1)_Y$ down to $U(1)_{em}$, just as the EW vacuum in the SM,
therefore preserving both CP and the electromagnetic symmetry.

But vacua with a spontaneous breaking of CP are also possible, and in fact their existence is the main
 reason the model was created by T.D. Lee~\cite{Lee:1973iz}. Such vacua occur when the doublets have
neutral vevs, but now, unlike eq.~\eqref{eq:vevn}, a relative complex phase between them appears, {\em i.e.}
the vevs are of the form
\be
\langle\Phi_1 \rangle_{CP} = {\displaystyle\frac{1}{\sqrt{2}}}\begin{pmatrix} 0 \\ \bar{v}_1 \end{pmatrix} \; , \;
\langle\Phi_2 \rangle_{CP} = {\displaystyle\frac{1}{\sqrt{2}}}\begin{pmatrix} 0 \\ \bar{v}_2\,\exp^{i\theta} \end{pmatrix}\,,
\label{eq:vevcp}
\ee
with $\theta \neq n\pi$, for any integer $n$. The complex phase induces spontaneous CP breaking and the
resulting scalar mass eigenstates have no definite CP quantum numbers -- they are neither CP-even nor
CP-odd. As a consequence, the neutral mass matrix in such minima is more complex than the analogous
matrix in normal vacua: in the latter, a $4\times 4$ matrix breaks into two $2\times 2$ blocks, one
having two non-zero eigenvalues, corresponding to the masses of the CP-even states $h$ and $H$, the
other having a zero eigenvalue (the Goldstone boson $G^0$) and the pseudoscalar mass of $A$; in the former
case, the $4\times 4$ matrix does not reduce to two blocks, it possesses a zero eigenvalue (again the
neutral Goldstone) and three eigenstates with interactions such that they are neither scalars nor
pseudoscalars.

Charge breaking vacua are also a possibility, where the upper components
of the doublets also acquire vevs, {\em i.e.} we will have
\be
\langle\Phi_1 \rangle_{CB} = {\displaystyle\frac{1}{\sqrt{2}}}\begin{pmatrix} 0 \\ v^\prime_1 \end{pmatrix} \; , \;
\langle\Phi_2 \rangle_{CB} = {\displaystyle\frac{1}{\sqrt{2}}}\begin{pmatrix} v^\prime_3 \\ v^\prime_2 \end{pmatrix}\,.
\label{eq:vevcb}
\ee
These minima, of course, are to be avoided at all costs -- the charged vev $v^\prime_3$ above will break the
electromagnetic symmetry and give the photon a mass. In the scalar mass matrix, the neutral (lower) components
of the doublets now appear mixed with the charged ones (upper), the resulting $8\times 8$ mass matrix having
a total of four zero eigenvalues -- corresponding to the expected four Goldstone bosons arising from the
full breaking of the gauge symmetry group.

The existence of a diverse number of minima in the potential raises the possibility of tunneling between
different vacua,
and certainly the hypothetical existence of, for instance, a CB minimum deeper than a EW or CP one,
could constitute a problem for the model. However, it has been shown
that if a normal minimum exists, any CP or charge breaking solutions of the minimisation equations
are necessarily saddle points
which lie above the normal minimum \cite{Ferreira:2004yd,Barroso:2005sm,Ivanov:2006yq,Ivanov:2007de}.
In fact, it was possible to show that the value of the potential at normal vacua ($V_N$), CP
stationary points ($V_{CP}$) or CB ones ($V_{CB}$) can be related to one another, for coexisting
tree-level stationary points of these types. The following formulae have been established:
\bea
V_{CB}\,-\, V_N &=& \left(\frac{m^2_{H^\pm}}{4 v^2}\right)_N\,
\left[(v_1 v^\prime_2 - v_2 v^\prime_1)^2 + v_1^2 {v^\prime_3}^2\right]
\label{eq:difvcb} \\
 & & \nonumber \\
V_{CP}\,-\, V_N &=& \left(\frac{m^2_A}{4 v^2}\right)_N\,
\left[(v_1 \bar{v}_2 \cos\theta - v_2 \bar{v}_1)^2 + v_1^2 \bar{v}_2^2 \sin^2\theta\right] \,,
\label{eq:difvcp}
\eea
with the vevs for each possible stationary points defined in eqs.~\eqref{eq:vevn}, ~\eqref{eq:vevcp}
and~\eqref{eq:vevcb}, and the subscript ``$N$" refers that the masses $m_{H^\pm}$, $m_A$ and the vev
$v$ are computed at the normal stationary point. The terms within the square brackets are
obviously positive thus, if $N$ is a minimum, its squared scalar masses will all be positive --
and hence these expressions show that $V_{CB}\,>\, V_N$ and $V_{CP}\,>\, V_N$ when $N$ is a
minimum. It is also easy to show that in that case both $CP$ and $CB$ stationary points
would necessarily be saddle points. Analogously, if the potential is such that a $CP$ ($CB$)
minimum occurs, any eventual normal or $CB$ ($CP$) stationary points would live above
the minimum and be saddle points. Thus tunneling to deeper minima of a different nature is
impossible in the 2HDM.

There is however another aspect of the 2HDM vacuum structure which sets it apart from the SM, to wit,
in certain situations the minimization conditions allow for several non-equivalent normal stationary
points \cite{Barroso:2007rr}. Therefore, already at tree-level, {\em there is the possibility of two (no more
than two) normal minima coexisting in the potential, at different depths}~\cite{Ivanov:2006yq,Ivanov:2007de}.
In other words, other than the normal vacuum with vevs given by eq.~\eqref{eq:vevn}, for which
one has $v_1^2 + v_2^2 = v^2 = (246$ GeV$)^2$, there may exist a second normal vacuum $N^\prime$,
with different vevs $\{v^\prime_1 , v^\prime_2\}$. For this second minimum of the potential, the sum of the squared
vevs takes a different value, smaller or larger than $(246$ GeV$)^2$.
The two minima are not degenerate, in fact they verify~\cite{Barroso:2012mj,Barroso:2013awa}
\be
V_{N^\prime} - V_{N}\,=\,\frac{1}{4}\,\left[\left(\frac{m^2_{H^\pm}}{v^2}\right)_{N}
- \left(\frac{m^2_{H^\pm}}{v^2}\right)_{N^\prime}\right]\,
(v_1 v^\prime_2 - v_2 v^\prime_1)^2\, ,
\label{eq:diffn}
\ee
where the quantity $\left(m^2_{H^\pm} / v^2\right)$ is evaluated at each of the minima, $N$ and
$N^\prime$. This raises the possibility that our vacuum, with $v =$ 246 GeV, is not the
deepest one -- there is nothing, in eq.~\eqref{eq:diffn}, which privileges the minimum $N$ over $N^\prime$,
unlike what happened in eqs.~\eqref{eq:difvcb} or~\eqref{eq:difvcp}. In fact, for certain regions of the
2HDM potential, $N^\prime$ may be found
to be the global minimum of the model - a minimum where the exact same symmetries have been
broken, but where {\em all elementary particles have different masses}. In that situation
our universe could tunnel to this deeper minimum, with obvious catastrophic consequences.

The conditions under which this rather intriguing possibility arises were established
in~\cite{Ivanov:2006yq,Ivanov:2007de,Barroso:2012mj,Barroso:2013awa}. Defining the quantity
\be
k = \sqrt[4]{\frac{\lambda_1}{\lambda_2}}\,,
\ee
the necessary (but not sufficient) conditions for the softly broken $\Z_2$ 2HDM potential to have two minima
are
\bea
m_{11}^2 + k^2\, m_{22}^2
&<& 0,
\label{eq:M0} \vspace{0.5cm}
\\
\sqrt[3]{x^2} + \sqrt[3]{y^2}
&\leq&  1,
\label{eq:astr}
\eea
where we have defined the variables $x$ and $y$ as
\bea
x
&=&
\frac{4\ k\ m_{12}^2}{
m_{11}^2 + k^2\, m_{22}^2}\,
\frac{\sqrt{\lambda_1 \lambda_2}}{
\lambda_{345} - \sqrt{\lambda_1 \lambda_2}}, \vspace{0.5cm}
\nonumber\\
y
&=&
\frac{m_{11}^2 - k^2\, m_{22}^2}{
m_{11}^2 + k^2\, m_{22}^2}\,
\frac{\sqrt{\lambda_1 \lambda_2} + \lambda_{345}}{
\sqrt{\lambda_1 \lambda_2} - \lambda_{345}}\,.
\label{eq:xy}
\eea
As shown in ref.~\cite{Barroso:2013awa}, {\em the EW vacuum ``N" (``our" minimum) is
the global, true, minimum of the theory if and only if $D\,>\,0$}, where the
discriminant $D$ is a quantity given by
\be
D \,=\, m^2_{12} (m^2_{11} - k^2 m^2_{22}) (\tan\beta - k)\, .
\label{eq:disc}
\ee
Notice how, remarkably, the value of $D$ can, in principle, be obtained by experiments performed
on ``our" minimum, without any knowledge of the existence of $N^\prime$.

Let us again recall (see the discussion following eq.~\eqref{eq:min}) that if the minimisation
conditions yield the solutions $N = \{v_1 , v_2\}$ and $N^\prime = \{v^\prime_1 , v^\prime_2\}$,
they also include other ``mirror" solutions, $\overline{N} = \{-v_1 , -v_2\}$ and
$\overline{N}^\prime = \{-v^\prime_1 , -v^\prime_2\}$. This is a trivial consequence of the fact that
the potential is invariant under a sign change of both doublets, $V(\Phi_1,\Phi_2) = V(-\Phi_1,-\Phi_2)$,
and {\em apparently } this has no physical consequences: the potential is degenerate at $N$ and $\overline{N}$
($N^\prime$ and $\overline{N}^\prime$), and physics at these two minima is entirely identical. No physical differences
whatsoever may arise from being at $N$ or $\overline{N}$ ($N^\prime$ or $\overline{N}^\prime$), because the only
difference between both minima is the overall sign of {\em both} fields -- no interference effects, for instance,
will be sensitive to the sign change. The SM minimum of the Higgs potential, of course, is also degenerate
with a continuum of other possible solutions -- recall the shape of the tree-level SM Higgs potential, where
infinitely many degenerate minima lie in a full circle. This is due to the fact that the SM minimum
is determined by the equation $\langle | \Phi| \rangle = v/\sqrt{2}$, which yields a continuum
of possible solutions, corresponding to different gauge choices for the Higgs doublet $\Phi$.
However, for the 2HDM potential, each of the minima $N$ and $N^\prime$ is not degenerate with a
continuum of other minima, but rather
with another separate isolated minimum, $\overline{N}$ and  $\overline{N}^\prime$ respectively.
We emphasize these seemingly trivial aspects
of the minimisation solutions because they may have dramatic consequences in the computation of
tunneling rates, as will be discussed below in section~\ref{sec:mindeg}.

The mere existence of a deeper minimum is however no valid reason to exclude the values of the
2HDM scalar parameters which produce it. In fact, should the lifetime of the false vacuum be larger
than the current age of the universe, such  situation, however strange, would be phenomenologically
acceptable. In~\cite{Barroso:2013awa} a quick estimate of the lifetime of 2HDM false vacua was
undertaken, but even then several shortcomings of the calculation were pointed out: the fact that
it was inspired by a single field tunneling computation, even though in the 2HDM the number of
fields varying from minimum to minimum is larger; the use of a ``thin-wall" approximation; the imposition
of a bounce trajectory necessarily passing by an intermediate saddle point. A thorough study of
the tunneling times between EW vacua is necessary to impose valid constraints on the 2HDM
parameter space, and we will undertake it in the next sections.

\section{Tunneling and bounces}
\label{sec:tun}

In the present section we briefly review the theoretical background
for the computation of the tunneling decay rate from a false vacuum
to a true vacuum, starting from the one field case studied
by Coleman and Callan \cite{Coleman:1977py,Callan:1977pt}, and extending then the result to the general $N$ field case.

Let us begin by considering the Euclidean action for a single component
real scalar field $\phi$:
\begin{equation}\label{maction}
S[\phi]=\int d^4x \left [ \frac 1 2 (\partial_\mu \phi)^2
+ V(\phi) \right ]\,,
\end{equation}
where $V(\phi)$ is a potential with a local minimum
(\emph{false vacuum}) at $\phi=\phi_{\rm fv}$, and
an absolute minimum (\emph{true vacuum}) at $\phi=\phi_{\rm tv}$. In order to calculate the
false vacuum lifetime we have to look
for the so called \emph{bounce solution} to the
Euclidean Euler-Lagrange equation that have
$O(4)$ symmetry and satisfy specific boundary
conditions\,\cite{Coleman:1977py}. Denoting by $r$ the euclidean radial coordinate,
$r=\sqrt{t_E^2+\boldsymbol x^2}$, where $t_E=-i t$ is the imaginary time, the action
(\ref{maction}) for $O(4)$ configurations takes the form
\begin{equation}\label{action}
S[\phi]=2 \pi^2 \int_0^\infty d r \ r^3
\left [ \frac 1 2 \left(\frac{d\phi}{dr}\right)^2 + V(\phi) \right ] \, ,
\end{equation}
while the equation of motion is
\begin{equation}\label{meq}
\frac{d^2 \phi}{dr^2} + \frac{3}{r} \, \frac{d \phi}{dr} = \frac{d V}{d \phi}\,,
\end{equation}
and the above mentioned
boundary conditions are:
\begin{equation}\label{mbc}
\phi(\infty)=\phi_{{\rm fv}} \qquad \mbox{and} \qquad\frac{d\phi(r)}{d r}\Bigg|_{r=0}=0\,.
\end{equation}

As the bounce solution $\phi_b(r)$ in the $r\to\infty$ limit goes to $\phi_{{\rm fv}}$, the action~\eqref{action} is infinite
when calculated at $\phi(r)=\phi_b(r)$. However, for the calculation of the tunneling time we need to subtract
to the bounce action the corresponding action calculated at the false vacuum $\phi_{{\rm fv}}$: $S[\phi_b]-S[\phi_{{\rm fv}}]$
(see below). Due to the asymptotic ($r\to\infty$) behaviour of $\phi_b(r)$~\cite{Dunne:2007rt,Dunne:2005rt},
the subtracted  action
\be
B\equiv S[\phi_b]-S[\phi_{{\rm fv}}]= 2\pi^2\int_0^\infty dr\,r^3\left[\frac12\left(\frac{d\phi_b}{dr}\right)^2+V(\phi_b)-V(\phi_{{\rm fv}})\right]
\label{eq:nova}
\ee
is finite.

Denoting by $B_K$ and $B_V$ the kinetic and the potential terms in
(\ref{eq:nova}), it is easily shown that $B_V=-1/2 \, B_K$, so that
$B=1/2 \, B_K= \frac{\pi^2}{2} \int_0^\infty d r \ r^3
\left(\frac{d\phi_b}{dr}\right)^2$. Integrating now by parts and using the equation of
motion (\ref{meq}), we get:

\begin{equation}\label{faction}
B=-\frac{\pi^2}{2} \int_0^\infty dr \, r^3 \, \frac{dV}{d\phi_b}\phi_b \, .
\end{equation}
The general formula for the decay rate $\Gamma$ of the false vacuum is
\cite{Coleman:1977py,Callan:1977pt}:
\be \label{gamma}
\Gamma =  D\, e^{-B}\,
\ee
and $B$ is usually called the {\it tunneling exponent}. The exponential of\, $-B$ \, gives the ``tree-level'' contribution
to the decay rate, while the prefactor $D$ contains the contributions from the quantum fluctuation
determinant, including those coming from the zero modes.

Denoting
 by $T_U$ the age of the universe, and approximating the prefactor as $D\simeq {T_U^3 \,\phi_b(0)^4} $
  \,\cite{Arnold:1991cv}, the tunneling rate $\Gamma$ in (\ref{gamma}) finally is:
\begin{equation} \label{tau_one}
\Gamma = \left[{T_U^3 \,\phi_b(0)^4}\right] \, e^{-B}\,,
\end{equation}
where $\phi_b(0)$ is the value of the bounce at $r=0$, which is the center of the bounce. The
tunneling time from the false to the true vacuum, {\em i.e.} the lifetime $\tau$ of the false vacuum, is
then given by the inverse of $\Gamma$: $\tau=\Gamma^{-1}$. Naturally, if the false vacuum can decay
towards more than one state, the tunneling rate $\Gamma$ is obtained by calculating the different
rates $\Gamma_i$, so that: $\Gamma =\sum_i \Gamma_i$, and again $\tau=\Gamma^{-1}$.

The extension to the case with $N$ real fields $\phi_i$, $i=1,...,N$ is straightforward.
If we denote the fields with $\boldsymbol \phi=(\phi_1, \phi_2, \dots , \phi_N)$ and the
potential as $V(\boldsymbol{\phi})$, the bounce configuration is a non-trivial solution of
the coupled system of $N$ ordinary differential equation:
\begin{equation}\label{bounceeq}
\frac{d^2 \phi_i}{dr^2}+\frac 3 r \frac{d \phi_i}{dr}=\frac{\partial V(\boldsymbol \phi)}{\partial \phi_i}
\end{equation}
with boundary conditions
\bea
\left. \frac{d \phi_i}{d r}\right|_{r=0} &=&0 \,,
\label{eq:bound1} \\
\lim_{r \to \infty} \phi_i &=&\phi_i^{{\rm fv}}\,,
\label{eq:bound2}
\eea
where $\phi_i =\phi_i^{{\rm fv}}$ are the values of the fields
$\phi_i$ at the false vacuum. Following the same steps that lead to Eq.\,(\ref{faction}), the
action calculated at the bounce solution $\boldsymbol {\phi}_b(r) = (\phi_1(r),\dots , \phi_N(r))_{bounce}$ for the $N$ field case takes the form:
\begin{equation}\label{finalaction}
B=-\frac{\pi^2}{2} \int_0^{\infty} dr \, r^3 \,
\Bigg[\frac{dV(\boldsymbol{\phi})}{d\phi_i}\phi_i\Bigg]_{\boldsymbol{\phi}_b} \, ,
\end{equation}
where a sum over $i$ is implied.

Apart from very simple cases, the system (\ref{bounceeq}) cannot be solved analytically and we have to
rely on numerical methods to evaluate the bounce configurations. To this end, we used the public
Wolfram Mathematica code developed in\,\cite{Masoumi:2016wot}. The latter solves the system (\ref{bounceeq}) with the help of a multiple shooting method, exploiting the asymptotic behavior of the bounce solution
for $r \to 0$ and $r \to \infty$ (that is known in both cases analytically).
Finally, the tunneling rate is given by:
\begin{equation} \label{tau}
\Gamma = {T_U^3 \left[ \sum_i \phi_i^2(0) \right]^2} \, e^{-B}\,.
\end{equation}

As for the one field case, if the false vacuum can decay towards more than one state, $\Gamma$ is obtained by calculating the different rates $\Gamma_i$\,: $\Gamma =\sum_i \Gamma_i$ and  $\tau=\Gamma^{-1}$.

For the 2HDM case, the two doublets have a total of eight real components, as seen in eq.~\eqref{eq:Phi}. Therefore,
in principle, the calculation of the bounce solution should involve all eight fields, which should contribute
to the tunneling time shown in eq.~\eqref{tau}. However, the gauge structure of the model allows a considerable
simplification of this procedure. In fact, since the model has a $SU(2)\times U(1)$ gauge invariance, we are
at liberty to choose a specific gauge, that allows us to remove several of the real components of the doublets.
This is a well-known feature of the 2HDM (see, for instance, the demonstration of this possibility in
section 5.8 of ref.~\cite{Branco:2011iw}) which, in passing, is also the reason why the most generic vacua
of the model can be cast into the form of eqs.~\eqref{eq:vevn},~\eqref{eq:vevcp} and~\eqref{eq:vevcb}. In the end,
we can choose to eliminate two of the upper components of the doublets (two charged fields) and one of the
imaginary components of
the lower part of the doublets, so that we are left with simplified doublets given by
\be
\Phi_1 \,=\,\frac{1}{\sqrt{2}}\,\left(
\begin{array}{c} 0 \\ \phi_1 \end{array}
\right)\;\;\; , \;\;\;
\Phi_2 \,=\,\frac{1}{\sqrt{2}}\,\left(
\begin{array}{c} \phi_4 + \mbox{i} \,\phi_5 \\ \phi_2 + \mbox{i} \,\phi_3 \end{array}
\right)\;,
\label{eq:phib}
\ee
where for convenience we have renamed the real component fields.

For the CP-conserving potential of eq.~\eqref{eq:pot} that we have been studying, the bounce
equation~\eqref{bounceeq} will allow a further simplification, involving only two of the
above component fields, namely $\phi_1$ and $\phi_2$. In fact, let us consider the derivatives of the
potential with respect to each of the $\phi_i$ that appear in the right-hand side of the bounce
equation~\eqref{bounceeq}. These are given by
\bea
\frac{\partial V}{\partial \phi_1} &=&  \frac{1}{2} \left[2 m_{11}^2 +  \lambda_1 \phi_1^2 +
\lambda_3 (\phi_2^2 + \phi_3^2 +\phi_4^2+\phi_5^2) +  \lambda_4 (\phi_2^2 + \phi_3^2 ) +
\lambda_5 (\phi_2^2 - \phi_3^2 )\right] \phi_1 - m_{12}^2 \phi_2
\label{eq:dphi1} \\
\frac{\partial V}{\partial \phi_2} &=&  \frac{1}{2} \left[2 m_{22}^2 +
\lambda_2 (\phi_2^2 + \phi_3^2 +\phi_4^2+\phi_5^2) +
(\lambda_3 +  \lambda_4 + \lambda_5) \phi_1^2  \right] \phi_2 - m_{12}^2 \phi_1 \\
\frac{\partial V}{\partial \phi_3} &=&  \frac{1}{2} \left[2 m_{22}^2 +
\lambda_2 (\phi_2^2 + \phi_3^2 +\phi_4^2+\phi_5^2) +
(\lambda_3 +  \lambda_4 - \lambda_5) \phi_1^2  \right] \phi_3 \\
\frac{\partial V}{\partial \phi_4} &=&  \frac{1}{2} \left[2 m_{22}^2 +
\lambda_2 (\phi_2^2 + \phi_3^2 +\phi_4^2+\phi_5^2) +
\lambda_3  \phi_1^2  \right]\phi_4 \,\\
\frac{\partial V}{\partial \phi_5} &=&  \frac{1}{2} \left[2 m_{22}^2 +
\lambda_2 (\phi_2^2 + \phi_3^2 +\phi_4^2 +\phi_5^2) +
\lambda_3  \phi_1^2  \right]\phi_5 \,.
\label{eq:dphi4}
\eea
Notice how in the three last equations the fields $\phi_3$, $\phi_4$ and $\phi_5$ factorize, and how that does not occur
for the derivatives of the potential with respect to $\phi_1$ and $\phi_2$. This leads to bounce equations for each
of the $\phi_i$ of the following form:
\bea
\frac{d^2 \phi_1}{dr^2}+\frac 3 r \frac{d \phi_1}{dr} &=& f_1(\phi_1,\dots\phi_5)\,\phi_1 \,-\,m_{12}^2 \phi_2
\label{eq:b1} \\
\frac{d^2 \phi_2}{dr^2}+\frac 3 r \frac{d \phi_2}{dr} &=& f_2(\phi_1,\dots\phi_5)\,\phi_2 \,-\,m_{12}^2 \phi_1
\label{eq:b2} \\
\frac{d^2 \phi_3}{dr^2}+\frac 3 r \frac{d \phi_3}{dr} &=& f_3(\phi_1,\dots\phi_5)\,\phi_3
\label{eq:b3} \\
\frac{d^2 \phi_4}{dr^2}+\frac 3 r \frac{d \phi_4}{dr} &=& f_4(\phi_1,\dots\phi_5)\,\phi_4\,
\\
\frac{d^2 \phi_5}{dr^2}+\frac 3 r \frac{d \phi_5}{dr} &=& f_5(\phi_1,\dots\phi_5)\,\phi_5\,,
\label{eq:b4}
\eea
where the functions $f_i$ can be read from eqs.~\eqref{eq:dphi1}--\eqref{eq:dphi4}. These equations
must be solved with the boundary conditions~\eqref{eq:bound1} and~\eqref{eq:bound2}. In our
case, for which both the true and false vacua of the CP conserving potential are themselves CP and charge conserving,
the boundary condition~\eqref{eq:bound2} always implies $\phi_3 (\infty)= \phi_4 (\infty)= \phi_5 (\infty)=0$ at any vacua.

We observe that
there is a fundamental difference between the bounce equations for $\{\phi_1,\phi_2\}$ and those for
$\{\phi_3,\phi_4,\phi_5\}$ -- namely, in the right-hand side of the latter equations the factorization of the fields
$\phi_3$, $\phi_4$ and $\phi_5$ implies that the trivial solutions $\phi_3 (r) = 0$, $\phi_4 (r) = 0$ and $\phi_5 (r) = 0$
exist. Moreover,
they respect the above-mentioned boundary conditions, and thus are acceptable bounce solutions.
On the contrary, in the right-hand side of the first two equations there is an extra term linear in the fields $\phi_1$
and $\phi_2$. And, though the trivial solutions $\phi_1 (r) = 0$ and $\phi_2 (r) = 0$ also satisfy
eqs.~\eqref{eq:b1} and~\eqref{eq:b2}, they do not comply with the boundary conditions at infinity for these two fields,
which are of the form $\phi_1 (\infty) = v_1$ and $\phi_2 (\infty) = v_2$ with non-zero values for the false vacua
vevs $v_1$ and $v_2$~\footnote{Notice how the soft breaking term $m^2_{12}$ in the potential prevents solutions of
the minimisation conditions of eq.~\eqref{eq:min} with any of the vevs equal to zero.}: thus they are not bounce solutions.

This strongly suggests that the bounce solutions connecting the true and false vacua
have the profiles $\phi_3 (r)$, $\phi_4 (r)$ and $\phi_5 (r)$ identically vanishing in the whole range for
$r$, from $0$
to $\infty$. This would imply that the original 2HDM 8-field bounce calculation reduces to a 2-field problem. In fact,
in all the
hundreds of thousands of cases that we have studied numerically (see section~\ref{sec:res}), we have always
verified that only $\phi_1(r)$ and $\phi_2(r)$ have non-trivial profiles, while $\phi_3(r)$, $\phi_4(r)$ and $\phi_5(r)$
always vanish~\footnote{Notice however that we do not possess a full analytical demonstration of this property.}.

This is not merely a mathematical property of the bounce equations~\eqref{bounceeq}, but rather it is dictated by
the physics of the model. To illustrate this point, let us consider for the moment the Complex 2HDM
(C2HDM)~\cite{Pilaftsis:1999qt,Ginzburg:2002wt,Khater:2003wq,ElKaffas:2007rq,ElKaffas:2006gdt,WahabElKaffas:2007xd,Osland:2008aw,
Grzadkowski:2009iz,Arhrib:2010ju,Barroso:2012wz}, where no CP symmetry is imposed on the potential of
eq.~\eqref{eq:pot2}. In this generalisation, both parameters $m^2_{12}$ and $\lambda_5$ can be complex although one
of these phases can always be absorbed into one of the fields. We are then left with a single complex parameter
in the potential, which we choose as the soft breaking term. Let us then write $m^2_{12} = |m^2_{12}|\exp^{\theta_{12}}$.
It is well known~\cite{Barroso:2007rr} that this potential may have coexisting minima as well -- now, however, there
is the possibility that in one of these minima the vevs of the doublets are real (as in eq.~\eqref{eq:vevn}) and in the
other the vevs have a relative complex phase (as in eq.~\eqref{eq:vevcp}). But since the potential explicitly breaks
the CP symmetry due to the presence of the phase $\theta_{12}$, {\em both} of these vacua are CP breaking, even if the vevs are real. For the C2HDM potential with complex $m^2_{12}$, then, the derivatives of the potential with respect to $\phi_i$
are slightly modified, and the bounce equations~\eqref{eq:b1}--\eqref{eq:b4} become
\bea
\frac{d^2 \phi_1}{dr^2}+\frac 3 r \frac{d \phi_1}{dr} &=& f_1(\phi_1,\dots\phi_5)\,\phi_1 \,-\,|m_{12}^2|
(\phi_2 \cos\theta_{12} - \phi_3 \sin\theta_{12})
\label{eq:bl1} \\
\frac{d^2 \phi_2}{dr^2}+\frac 3 r \frac{d \phi_2}{dr} &=& f_2(\phi_1,\dots\phi_5)\,\phi_2 \,-\,\phi_1 |m_{12}^2|
 \cos\theta_{12}
\label{eq:bl2} \\
\frac{d^2 \phi_3}{dr^2}+\frac 3 r \frac{d \phi_3}{dr} &=& f_3(\phi_1,\dots\phi_5)\,\phi_3  +
\phi_1 |m_{12}^2|\sin\theta_{12}
\label{eq:bl3} \\
\frac{d^2 \phi_4}{dr^2}+\frac 3 r \frac{d \phi_4}{dr} &=& f_4(\phi_1,\dots\phi_5)\,\phi_4
\\
\frac{d^2 \phi_5}{dr^2}+\frac 3 r \frac{d \phi_5}{dr} &=& f_5(\phi_1,\dots\phi_5)\,\phi_5\,.
\label{eq:bl4}
\eea
Comparing the system of equations \eqref{eq:b1}--\eqref{eq:b4} with the corresponding system
\eqref{eq:bl1}--\eqref{eq:bl4} we observe that, while the two last equations remain unchanged, the
right hand side of the third equation contains an additional term that does not factorize $\phi_3$
(further, the non-factorized terms in eqs.\eqref{eq:bl1}--\eqref{eq:bl2} have also changed). Thus,
we no longer expect a trivial profile for the bounce solution $\phi_3(r)$. Clearly, the appearance
of the additional term in the bounce equation for $\phi_3$, which we recall is the complex neutral component
of the second doublet, depends on the presence of the explicitly CP breaking phase $\theta_{12}$:
the different physics described by the C2HDM induces a different structure in the bounce equations.
\begin{figure}[t!]
	\centering
	\includegraphics[width=0.6\linewidth]{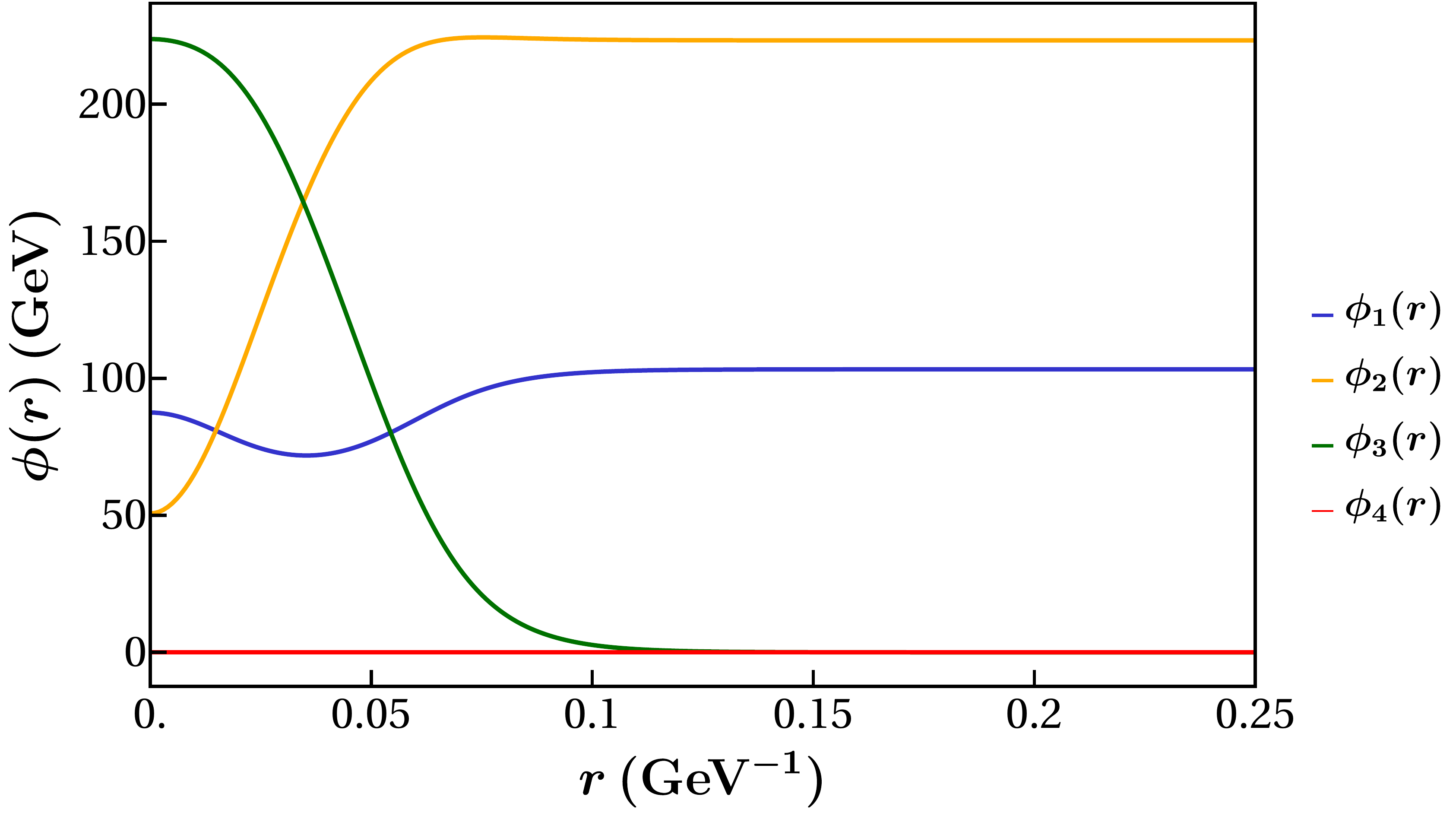}
	\caption{Bounce solution ($\phi_1$, $\phi_2$, $\phi_3$, $\phi_4$) for the C2HDM. The
fifth bounce for $\phi_5$ is identical to that for $\phi_4$.}
	\label{fig:bounce}
\end{figure}
At this point, we consider many different choices for the parameters of the C2HDM potential in which coexisting
minima occur. These points are chosen such that the false vacuum has real vevs, while the vevs of the true minimum
have a relative
complex phase. Computing the bounce solution for this parameter space, our expectation for the bounce profiles
is fully confirmed: for this new model, $\phi_1$, $\phi_2$ {\em and} $\phi_3$ are non trivial profiles,
while $\phi_4$ and $\phi_5$ vanish
as before. We see a particular example of this behaviour in fig.~\ref{fig:bounce}, where we plot
the different fields of the bounce solution $\phi_i$ as a function of $r$. We remind the reader that in this plot the fields
tend at $r \rightarrow \infty$ to the false vacua vevs -- and thus $\phi_3$ in that limit vanishes, as expected.
As opposed to what happened in the CP conserving potential, however, $\phi_3(r)$ is no longer vanishing everywhere.
In particular, we observe that at $r = 0$ it is taking a non-zero value, thus contributing (as well as the other non-zero
components of the bounce) to the evaluation of the tunneling time in eq.~\eqref{tau}.

It is worth stressing at this stage that this change in the behaviour of $\phi_3(r)$ is due to the different
physics described by the two potentials. Only due to the explicit CP violation of the C2HDM can $\phi_3$ have a
non-trivial profile, while explicit CP conservation forces this component of the bounce to vanish for all
values of $r$. Further, notice that for both potentials $\phi_4$ and $\phi_5$ are always vanishing -- which is to be expected
on physical grounds, since no charge breaking can occur in either of the models when a normal minimum exists.

\section{Tunneling to degenerate vacua}
\label{sec:mindeg}

At this stage, and before we embark on scans of the 2HDM parameter space, let us discuss a novel
aspect of the tunneling calculations which arise in this model. As we have emphasised previously,
if the minimisations conditions~\eqref{eq:min} have a solution of the form $\{ v_1\,,\,v_2 \}$, they
also include the solutions  $\{-v_1\,,\,-v_2\}$. The same happens for the second,
non degenerate minimum $N^\prime$, which corresponds to vevs of the form $\{v^\prime_1\,,\,v^\prime_2\}$.

Let now $N \equiv \{v_1\,,\,v_2\}$ and $\overline{N}\equiv \{-v_1\,,\,-v_2\}$ be the false vacua of the model,
and assume that ``our" vacuum corresponds to $N$. The universe may now tunnel to {\em TWO} degenerate
true vacua, $N^\prime = \{v^\prime_1\,,\,v^\prime_2\}$ and $\overline{N}^\prime = \{-v^\prime_1\,,\,-v^\prime_2\}$.
Since $N^\prime$ and $\overline{N}^\prime$ describe exactly the same physics, one could expect that there
would be absolutely no difference between the tunneling rates from $N$ to either $N^\prime$ or $\overline{N}^\prime$.
This, remarkably, is not the case.

In order to understand this critical point, let us consider a specific example, for which the parameters of the
2HDM potential~\eqref{eq:pot2} are chosen to be
\bea
m^2_{11} &=&  -23519.8\;\;\;,\;\;\; m^2_{22} = -10249.6;\;\;,\;\;\; m^2_{12} = -6145.98\;\;\;(\mbox{GeV}^2) 
\nonumber \\
\lambda_1 &=& 4.59143\;\;,\;\; \lambda_2 = 0.388928\;\;,\;\; \lambda_3 = 1.79703\;\;,\;\;
 \lambda_4 =  -1.80544\;\;,\;\; \lambda_5 = -0.481738\,.
 \label{eq:param}
\eea
This choice of parameters yields a maximum $M$ at field values $M \equiv \{\phi_1\,,\,\phi_2\} = \{0\,,\,0\}$ and
the following minima (all vevs in GeV),
\bea
N\equiv \{ 97.3767\,,\, 225.907\} & , & \overline{N} \equiv \{ -97.3767\,,\, -225.907\} \nonumber \\
N^\prime \equiv \{ 162.491 \,,\, -319.463 \} & , & \overline{N}^\prime \equiv \{ -162.491 \,,\, 319.463\} \,.
\eea
We also have two couples of saddle points,
\bea
S_1 \equiv \{ 43.6574\,,\, 221.06 \} & , & \overline{S}_1 \equiv \{ -43.6574\,,\, -221.06\} \nonumber \\
S_2 \equiv \{ 95.5578\,,\, 48.8458 \} & , & \overline{S}_2 \equiv \{ -95.5578\,,\, -48.8458\}\,.
\eea

If we now calculate the bounce solutions for the transitions from $N$ to $N^\prime$ and from $N$ to $\overline{N}^\prime$,
and assuming for the sake of argument that only one of these transitions was possible, we would obtain
the following tunneling times (see eq.~\eqref{tau}),
\bea
\tau(N\rightarrow N^\prime) & \simeq & 8\times 10^{2131}\,T_U\,, \nonumber \\
\tau(N\rightarrow \overline{N}^\prime) & \simeq & 2\times 10^{-113}\,T_U\,,
\label{eq:tunt}
\eea
where $T_U$ is the current age of the universe~\footnote{The factors multiplying $T_U$ in eq.~\eqref{eq:tunt} are either
ridiculously low or ridiculously large. Recall, however, that they stem from the exponential of the action~\eqref{faction},
which is {\em quartic} in the fields. Since in the bounce solution the fields acquire values of the order of the
hundreds of GeV, the action will be either positive or negative, but always include terms of the order of
$10^8$ GeV$^4$, which leads to such stunning numbers as those in eq.~\eqref{eq:tunt}.}.
If one were to only consider the transition $N\rightarrow N^\prime$
one would conclude that the false vacuum $N$ was absolutely stable -- whereas the second
transition, $N\rightarrow \overline{N}^\prime$, instead shows $N$ to be incredibly unstable, having
decayed to $\overline{N}^\prime$ almost immediately after the Big Bang. The discrepancy between the tunneling times for both
transitions is astonishing, all the more so because the lower minima $N^\prime$ and $\overline{N}^\prime$
are degenerate and describe exactly the same physics! Thus one might na\"{\i}vely expect that there
should be no difference in the tunneling rate from $N$ to either of them -- after all the difference
in the value of the potential between $N$ and $N^\prime$ or between $N$ and $\overline{N}^\prime$ is
exactly the same, and given by eq.~\eqref{eq:diffn}.
\begin{figure}[t!]
  \centering
  \includegraphics[width=0.56\linewidth]{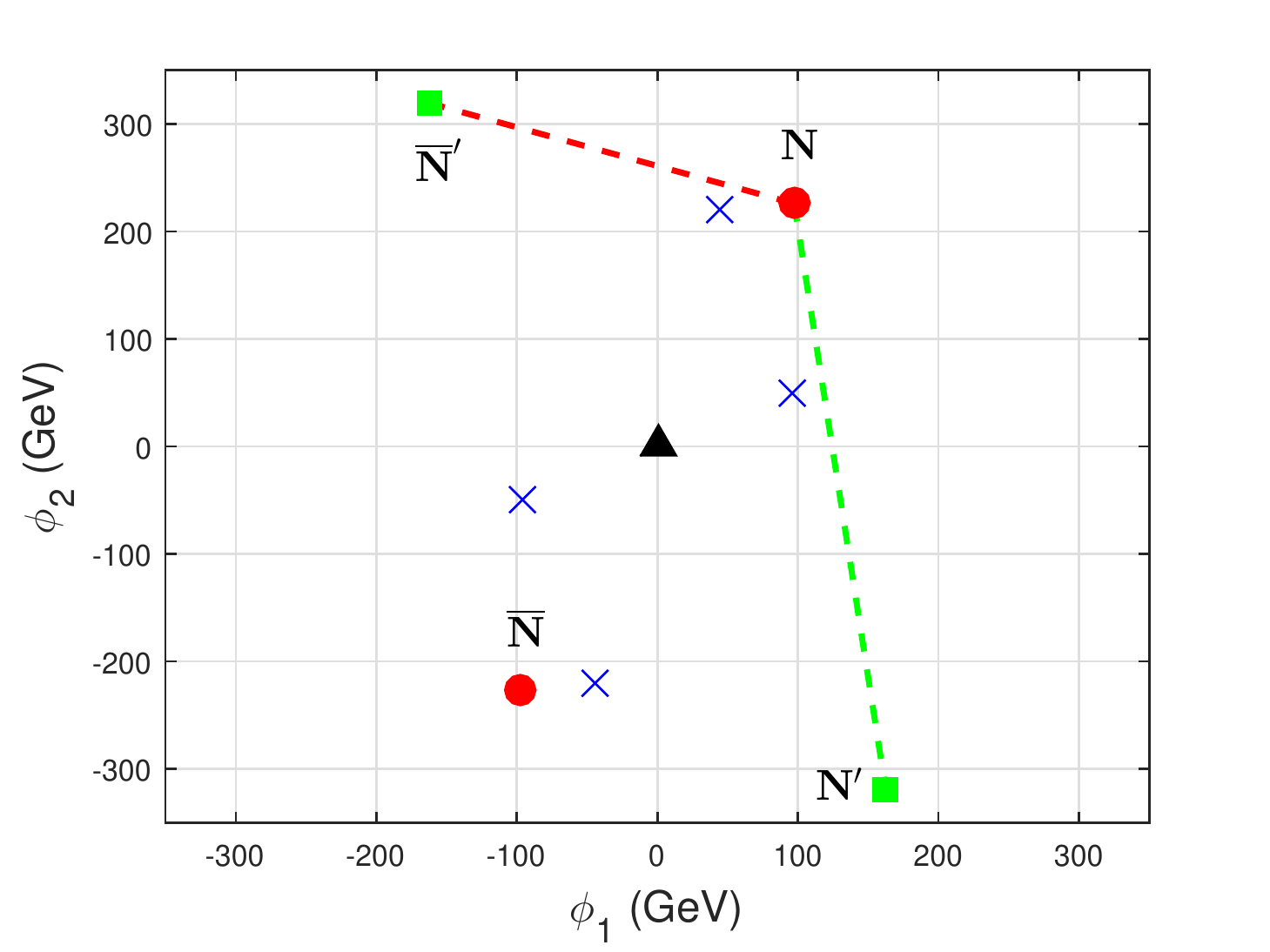}
  \caption{Location of all extrema of the 2HDM potential for the choice of parameters in~\eqref{eq:param}. Saddle
  points are marked with ``$\times$", the maximum of the potential, at $(0,0)$, with a black triangle. The false minima
  are marked with red circles, the true ones with green squares. The lines connecting $N$ to
  $N^\prime$ and $\overline{N}^\prime$ illustrate how different the paths between these minima
  may be.}
 \label{fig:lands}
\end{figure}
How can such a difference in behaviour be
explained? The fundamental reason is extremely simple to understand, and lies in the landscape of minima and
saddle points yielding very different possible paths for tunneling between $N$ and $N^\prime$ or $\overline{N}^\prime$.
This may be seen in fig.~\ref{fig:lands}, where we illustrate, in the $\{\phi_1\,,\,\phi_2\}$ plane, the
locations of all extrema of the potential listed above. Notice how $N$ is {\em not} equally distant from
$N^\prime$ and $\overline{N}^\prime$; notice also, and perhaps even more importantly, that the path from
$N$ to both of the lower minima passes close to a different landscape of saddle points -- whereas from
$N$ to $\overline{N}^\prime$ there is a saddle point almost at the beginning, to $N^\prime$ the first
saddle point is further away. Also, the steepest descent from $N$ to $N^\prime$ is possibly ``deviated" by the several
remaining saddle points and the maximum along the way, which would explain the much larger tunneling time found, whereas the
\begin{figure}[t!]
\begin{tabular}{cc}
\includegraphics[width=0.47\linewidth]{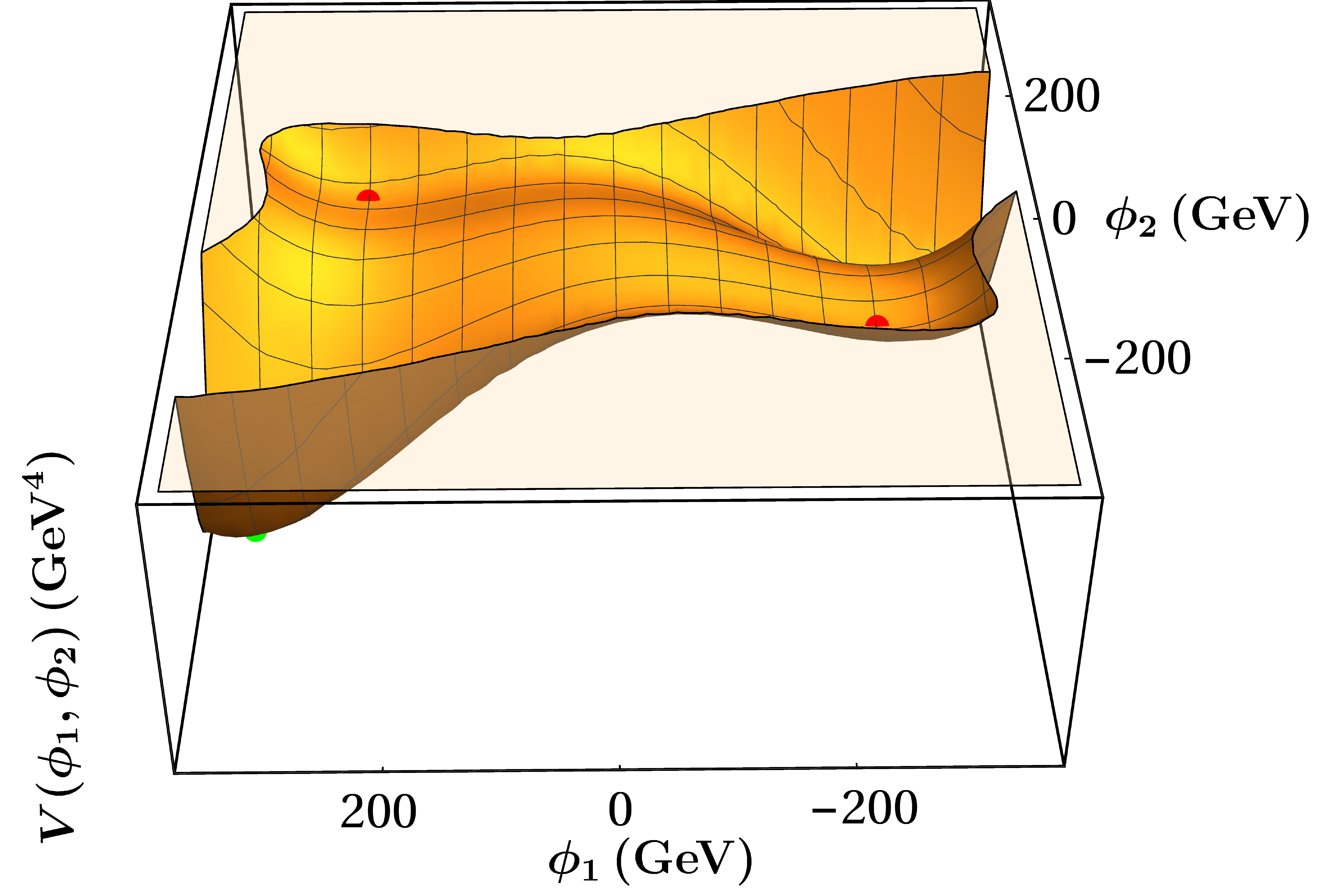}&
\includegraphics[width=0.47\linewidth]{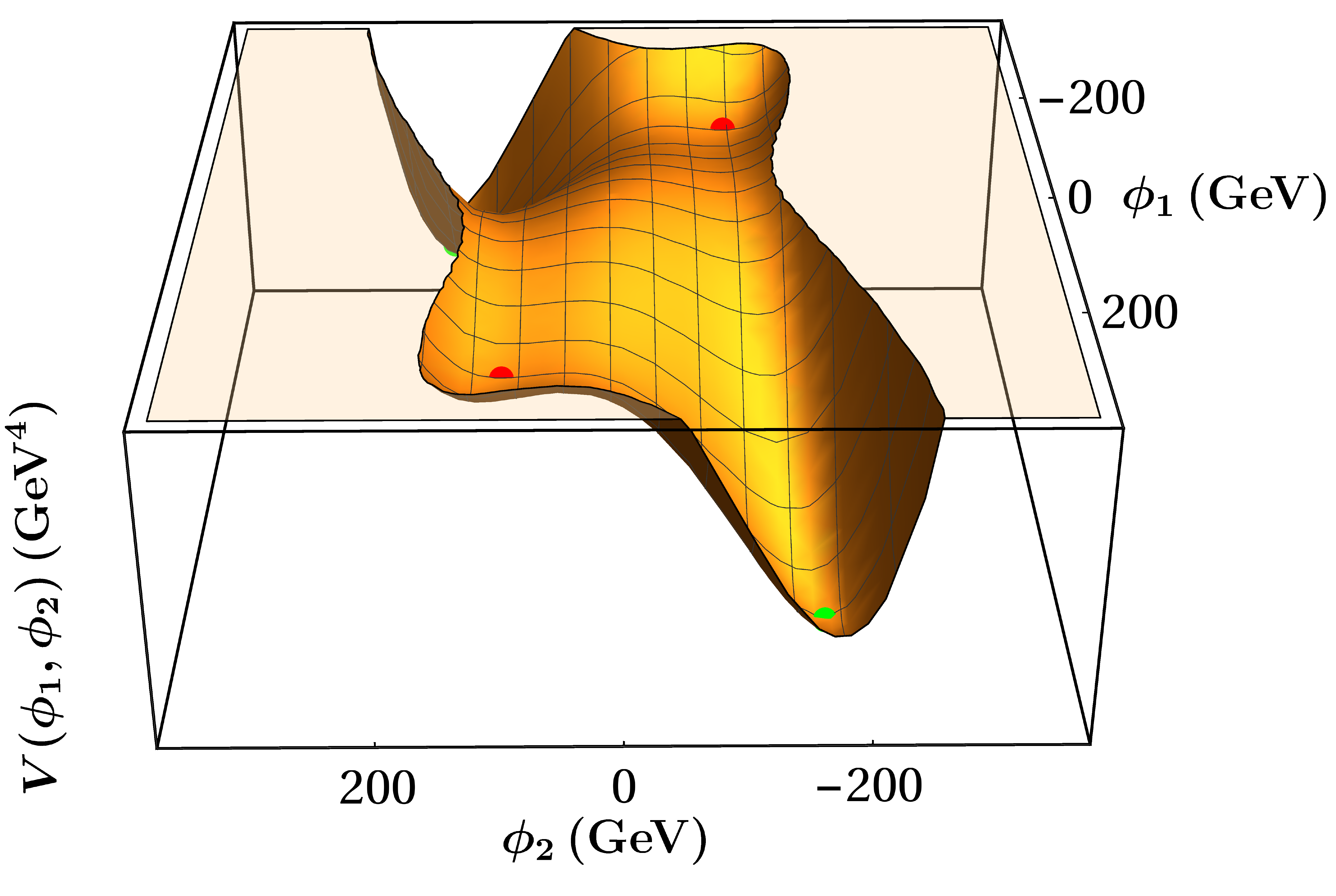}
\end{tabular}
\caption{Left panel: Plot of the potential $V(\phi_1,\phi_2)$ of eq.~\eqref{eq:pot2} for the parameters given of
eq.~\eqref{eq:param}. Right panel: The same potential rotated anticlockwise by 90 degrees. The left panel better
shows the decay path from $N$ to $N^\prime$; the right panel  from $N$ to $\overline{N}^\prime$. The path connecting
$N$ while $N^\prime$ is longer than the path connecting $N$ with $\overline{N}^\prime$. False minima marked in
red, true ones in green.
}
\label{fig:val}
\end{figure}
path to $\overline{N}^\prime$ seems much more ``direct". To further drive in this point, consider fig.~\ref{fig:val},
where we show 3D plots illustrating the shape of the potential along the (seemingly) shortest path from $N$ to both
$N^\prime$ and $\overline{N}^\prime$ -- these images show that, even though the difference in depth
of the potential is exactly the same between $N$ and $N^\prime$ or between $N$ and $\overline{N}^\prime$, it is
quite clearly easier for the latter transition to occur than the former. And in fact the bounce solutions
obtained in the transition from $N$ to $N^\prime$ (which we now call ``$B1$") and from $N$ to $\overline{N}^\prime$,
(``$B2$") are quite different, as can be appreciated in fig.~\ref{fig:bounced}.
\begin{figure}[t!]
  \centering
  \includegraphics[width=0.6\linewidth]{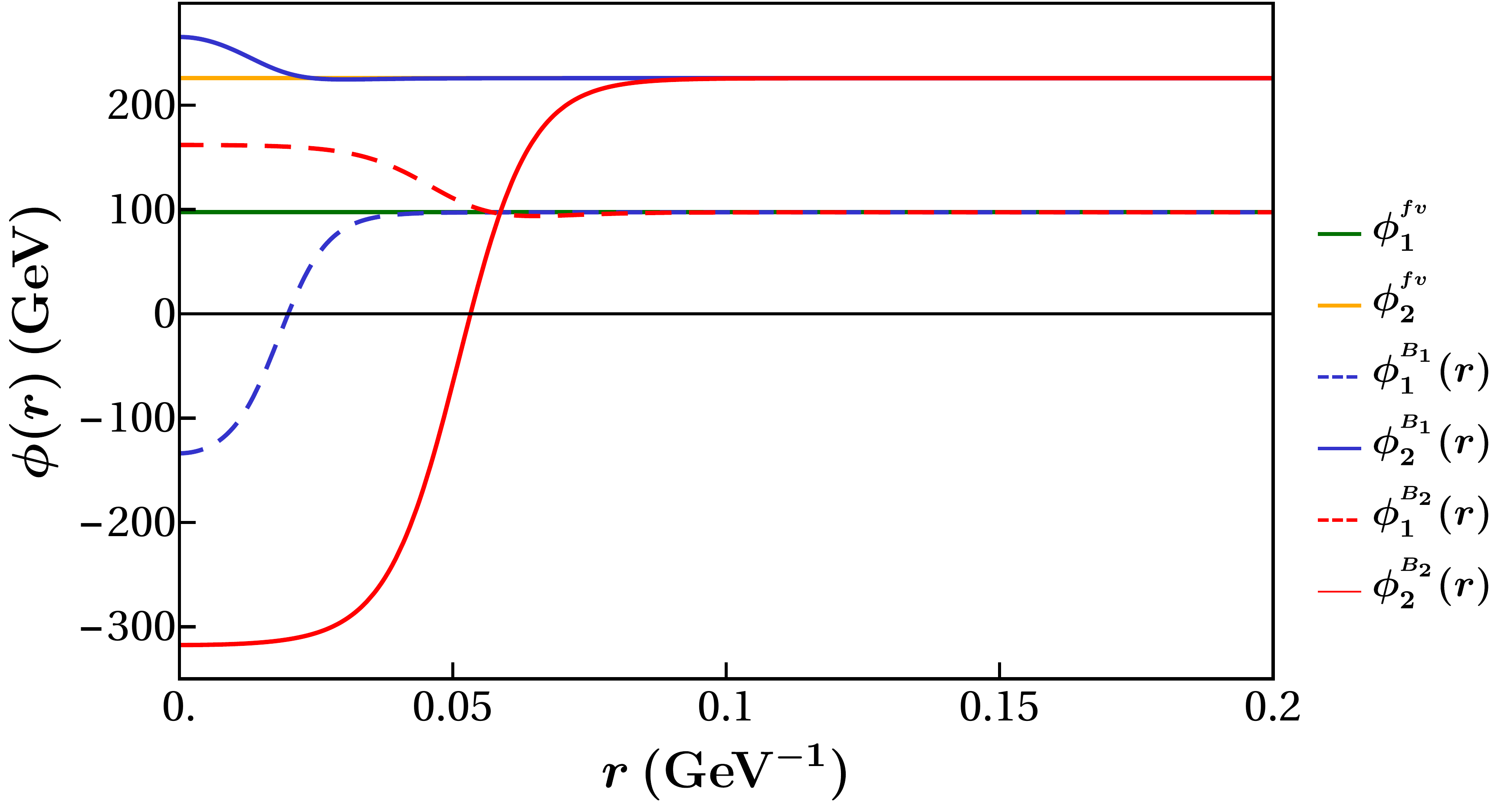}
  \caption{Bounce solutions for fields $\phi_1$ and $\phi_2$ for the transitions from $N$ to $N^\prime$
  ($B1$) and from $N$ to $\overline{N}^\prime$ ($B2$). In both cases the fields $\phi_i$ tend to the same values
  at $r\rightarrow \infty$, {\em i.e.} the values of the fields at the false vacuum $N$, $\phi_1^{{\rm fv}}$ e $\phi_2^{{\rm fv}}$.
  But at $r = 0$ the fields assume different values, close to the vevs at each of the degenerate true vacua.
  }
 \label{fig:bounced}
\end{figure}
In this plot we present the evolution with the radial coordinate $r$ (introduced in section~\ref{sec:tun})
of the two bounce profiles for the fields $\phi_1$ and $\phi_2$ found for the specific example we have been
considering. Notice how the solutions, $B1$ and $B2$, converge for large values of $r$ to the same values --
which are the values of the vevs at the false vacuum $N$, as was to be expected. However, the values of
the fields $\phi_i$ at $r = 0$ diverge significantly, assuming even opposite signs.
Recall that at $r = 0$ the bounce solution is found for values of the
fields ``close" to the true vacuum of the theory. Hence we find that, for the bounce $B1$, $\phi_1$ assumes a large
negative value, $\simeq -130$ GeV and $\phi_2$ a large positive one, $\simeq 260$ GeV -- notice how these values
for the bounce are close to the vevs of the true vacuum $\overline{N}^\prime$ ($\sim$ -162, $\sim$ 320 GeV).
Likewise, the values found for the bounce solution $B2$ are close to the vevs found for the other lower vacuum,
$N^\prime$. Thus, despite the fact that both $N^\prime$ and $\overline{N}^\prime$ are degenerate and at the same
relative depth to $N$, the bounce solutions for the two possible transitions are very different, and in fact lead
to very different values for the bounce action $S[\phi_b]$ from eq.~\eqref{faction} -- and hence to the two
extremely different lifetime values found.

If the potential has, from $N$, two possible ``decay channels", then its decay rate, $\Gamma$, will be given by
\be
\Gamma \,=\, \Gamma(N\rightarrow N^\prime) \,+\,\Gamma(N\rightarrow \overline{N}^\prime)\,=\,
\frac{1}{\tau(N\rightarrow N^\prime)} \,+\,\frac{1}{\tau(N\rightarrow \overline{N}^\prime)}\,
\ee
with the ``partial" tunneling times from eq.\eqref{eq:tunt}. Thus, the lifetime $\tau$ of the false vacuum
$N$ will obviously be
\be
\tau\,=\,\frac{1}{\Gamma}\,=\,\left(\frac{1}{\tau(N\rightarrow N^\prime)} \,+\,
\frac{1}{\tau(N\rightarrow \overline{N}^\prime)}\right)^{-1}\simeq\,2\times 10^{-113}\,T_U\,
\ee
where in analogy with nuclear decays, when one of the decay channels is much
faster than the other, it dominates over the total lifetime. The conclusion to draw from this particular example
is simple: both degenerate lower vacua must be considered for the calculation of the tunneling time, and the
stability of the false vacuum may depend crucially on which of the true vacua it is decaying into.
We have verified that differences in tunneling times to true degenerate vacua can be as extreme as
those presented in eq.~\eqref{eq:tunt} for many choices of parameters in the potential, though not always.
For many other regions of parameter space, though the two possible decay rates may differ, they do not
affect qualitatively the overall stability of the false vacuum. Meaning, in many cases, if the tunneling
time to one of the lower vacua is, say, much larger (smaller) than $T_U$, the other tunneling time, while
possibly very different, will also be much larger (smaller) than $T_U$. Nonetheless, as we will shortly see,
for certain regions the computation of $\tau$ taking into account the existence of both possible true vacua
increased the number of dangerous false vacua by as much as 50\%.

\section{2HDM Numerical Scans}
\label{sec:res}

The physics arguments of section~\ref{sec:tun} show that the tunneling rate calculation can be
reduced, for the CP conserving potential of eq.~\eqref{eq:pot2}, to a two-field problem. Nonetheless
we performed extensive numerical checks, comparing eight-field calculations with two-field ones, and
no differences were ever found. Also, in section~\ref{sec:mindeg} we have shown the importance
of computing the tunneling rates to {\em both} degenerate true vacua. Armed with these two important theoretical
insights, we can proceed to an extensive scan of the 2HDM parameter
space. Our goal is to ascertain how much of that parameter space should be avoided due to
tunneling times shorter than the age of the universe.

We have chosen to work in models type I and II (for the remaining types of Yukawa interactions
the conclusions reached would certainly be very similar). All parameter scans presented in this
section are such that:
\begin{itemize}
\item They include at least one (CP conserving) minimum with $v = 246$ GeV and $m_h = 125$ GeV.
\item All theoretical and experimental results mentioned in section~\ref{sec:exp} are
satisfied. In particular, we demanded that all $\mu_X$ ratios (defined in eq.~\eqref{eq:muX})
be within 30\% of their expected SM value of 1.
\item $1 \leq \tan\beta \leq 30$ and $-\pi/2 \leq \alpha \leq \pi/2$.
\item The mass of the heavier CP-even scalar $H$ is chosen in the interval between
130 and 700 GeV. The mass of the pseudoscalar $A$ is chosen between 100 and 700 GeV.
For the charged mass, its lower bound is 100 GeV for model type I and 580 GeV for model
type II (the difference due to flavour physics constraints described in
section~\ref{sec:exp}). The upper bound for the charged mass is again 700 GeV.
\item The soft breaking parameter $m^2_{12}$ is taken with both signs, and magnitude
below roughly (500 GeV)$^2$.
\end{itemize}
These parameter scans are not meant to be exhaustive -- we merely wish to sample representative
regions of the 2HDM parameter space to illustrate the possible impact that tunneling times to
deeper vacua lower than the age of the universe may have. We now consider different scenarios.

\subsection{General scans for type I and II models}

To illustrate the possible relevance of false vacua exclusion (due to low tunneling times)
in general scans of parameter space, we generated large datasets (over 100000 different combinations
of parameters) for models type I and II.
\begin{figure}[ht]
  \centering
  \includegraphics[height=6cm,angle=0]{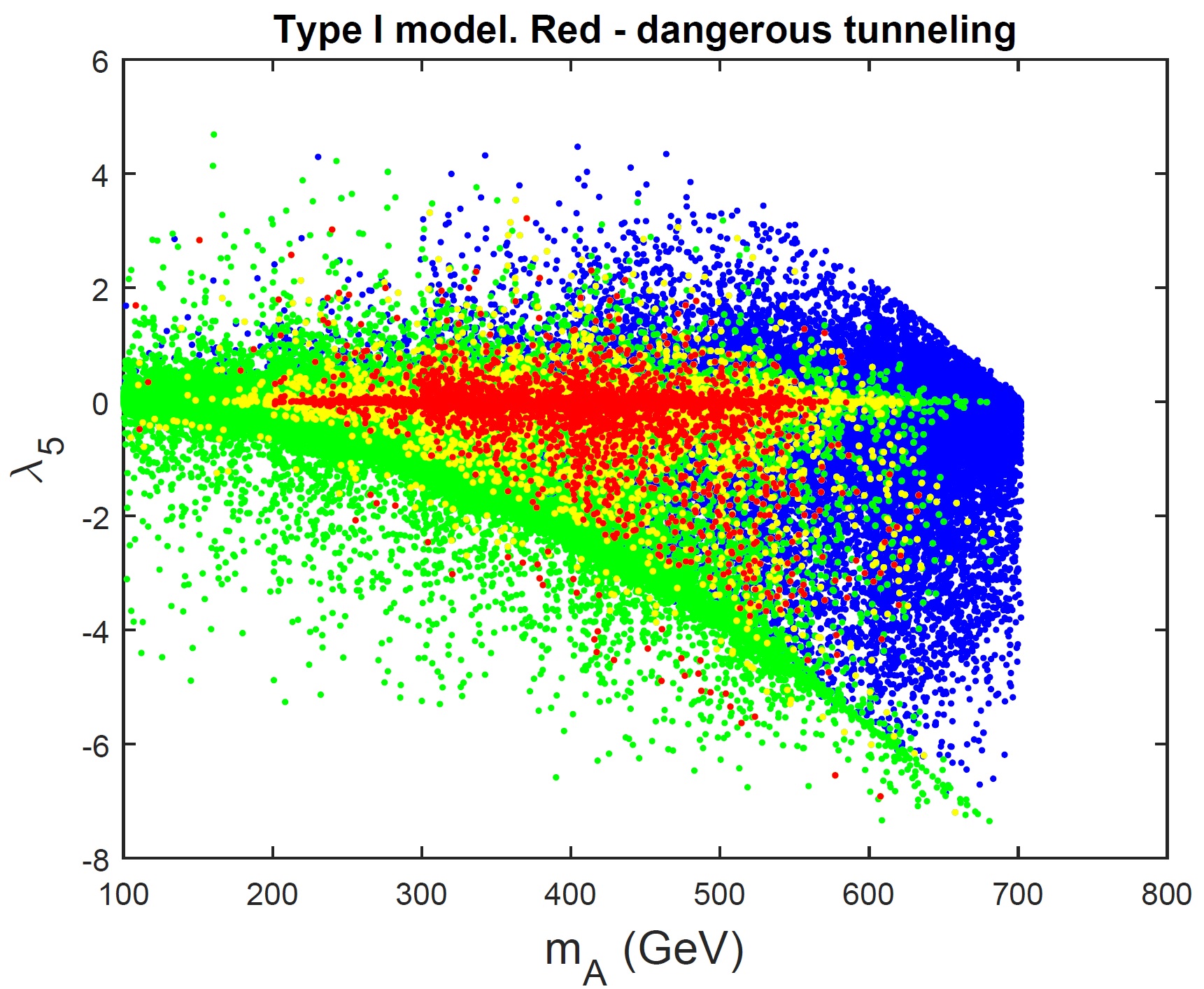}
  \includegraphics[height=6cm,angle=0]{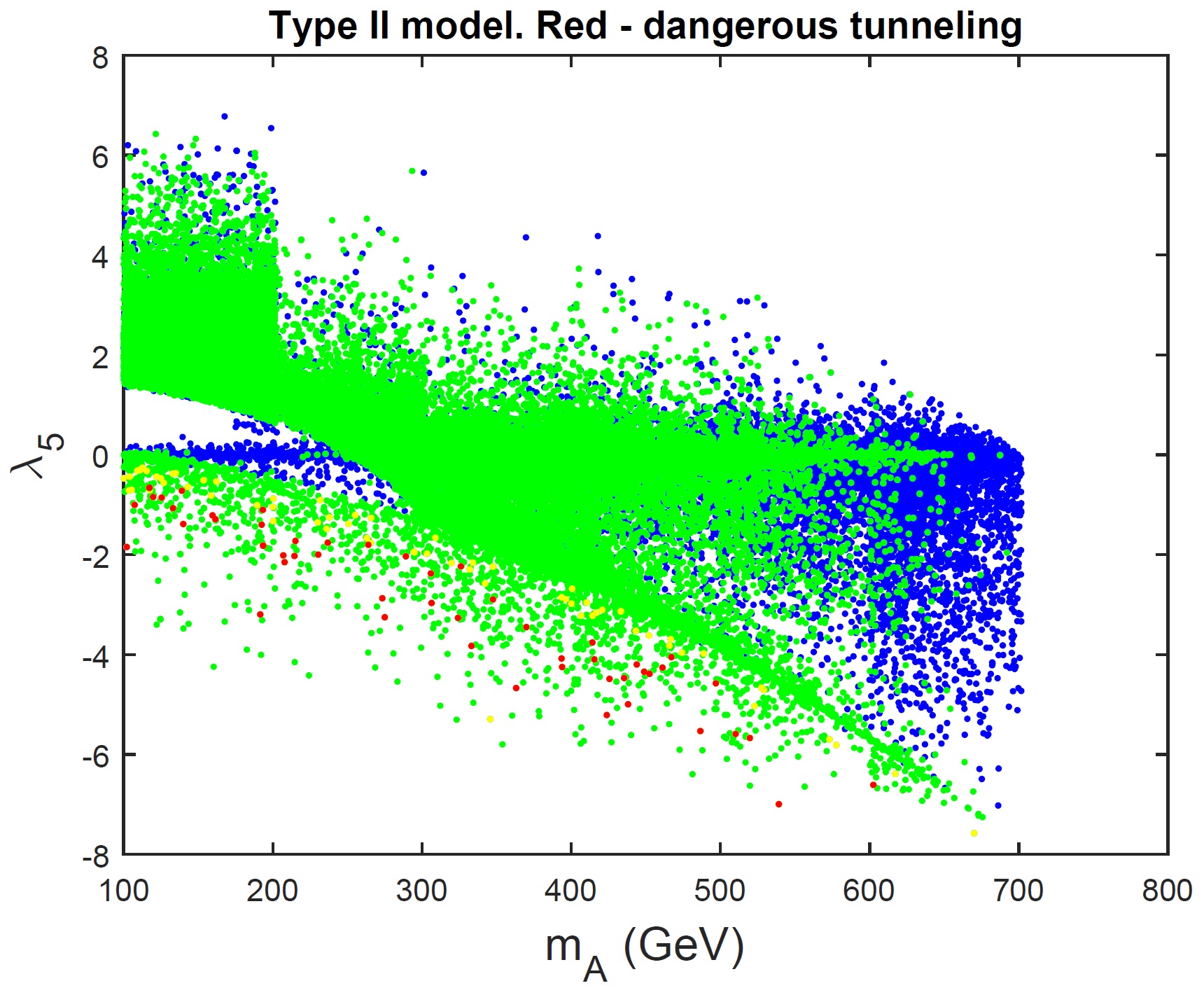}
  \caption{Scatter plot of $\lambda_5$  {\em vs.} $m_A$
  for  general scan on the parameter space of a type I (left) and type II (right) 2HDM.
  In blue, all points generated which conform to
  theoretical and experimental constraints; in green, the subset of those for which two
  normal vacua are possible; in yellow, the subset of those for which $D<0$ and thus the EW
  vacuum may be the false one; and in red, those points for which the tunneling time to the
  true vacuum is smaller than the age of the universe.}
 \label{fig:types}
\end{figure}

In fig.~\ref{fig:types} we show the result of our analysis, by plotting the values of $\lambda_5$
{\em vs} the pseudoscalar mass $m_A$. The colour code in these plots is such that:
\begin{itemize}
\item In blue we present all points generated which satisfy the theoretical and experimental
constraints explained above. Notice that other colours are superimposed on top of the
blue points. Or, in other words, the green, yellow, red points are a subset of the blue ones.
\item The green points correspond to the subset of the blue ones for which the two CP-conserving
minima conditions of eqs.~\eqref{eq:M0} and~\eqref{eq:astr} are satisfied. Recall that those conditions
are necessary ones, but not sufficient, and therefore not all green points will truly correspond
to the existence of two minima -- in fact, that happens typically for only half of these points.
Notice the disproportion in size of the green region compared to the blue one -- dual minima in
the 2HDM potential are, in general, a rare occurrence.
\item In yellow we show the subset of the green points for which the discriminant $D$ from eq.~\eqref{eq:disc}
is negative -- that is, the points for which, if two minima exist, ``our" electroweak vacuum with $v = 246$ GeV
will not be the deeper one~\footnote{See also~\cite{Xu:2017vpq}.}.
\item Finally in red, the subset of the yellow points for which: (a) two minima exist, (b) ``our" electroweak
vacuum is not the global minimum and (c) the tunneling time from ``our" vacuum to the deeper true vacuum
is less than the age of the universe.
\end{itemize}
The visible blue points in these figures are clearly safe combinations of parameters, for which the EW vacuum
is not only safe but also unique.
Several comments are in order to better interpret these plots. First, please take into account the fact that
these plots are dense in each of the colours -- in other words, in the middle of the green, or yellow, or red
points there are blue ones. Thus the red regions are not wholly excluded -- though dangerous tunneling times
seem to be found for specific areas in the $m_A$--$\lambda_5$ plane, those areas will in general also include
perfectly acceptable blue (green, yellow) points for which there might not even be two minima. Second, there is no
obvious pattern to the green, yellow or red regions -- the existence of two minima, and dangerous tunneling
times for the acceptable EW vacuum, depends on non-trivial relations between the potential's
parameters, which are difficult to visualize in this 2-dimensional slice of what is in truth an 8-dimensional
parameter space. Third, in general it seems easier to find two minima (and dangerous short-lived vacua) in
model type I than in II. This is a consequence of the hard bound on the charged Higgs mass in model type II
which arises from $b\rightarrow s\gamma$ constraints -- this bound tends to privilege higher, positive, values of
$m^2_{12}$, for which the discriminant $D$ is usually found to be positive (and thus ``our" EW vacuum is
the global minimum of the model).

To illustrate the frequency with which dangerous vacua are found in this blind scan, consider the results
shown above for type I: the total number of generated (blue) points conforming to all theory and
experimental constraints was above 120000; of these, roughly 21500 (green) points were found which
might have two minima (satisfying eqs.~\eqref{eq:M0} and~\eqref{eq:astr}) -- in fact, of those, two minima
were found only for over 11000 points. The (yellow) points with $D<0$, with possible local EW vacua with
$v= 246$ GeV, totalled almost 9500, and out of these over 4200 were found for which the tunneling time
to the global ($v \neq 246$ GeV) minimum is inferior to $T_U$. Thus the percentage of points of the
initial parameter space excluded on tunneling time arguments is about 3.5\%. For model type II, a similar
accounting yields a percentage of excluded points of roughly 0.2\%.

The distribution of dangerous (red) points in fig.~\ref{fig:types} is clearly not homogenous, and the percentages
of excluded points found in the previous paragraph are therefore not to be interpreted as, for instance,
3.5\% of type I parameter space being ruled out on low tunneling times grounds. In fact, while certain
regions of 2HDM parameter space are completely safe (the blue points visible in fig.~\ref{fig:types}, for
instance), others may yield a far larger percentage of dangerous minima than the numbers quoted above.
To illustrate this let us now consider a few benchmark scenarios.

\subsection{First benchmark scenarios: safe regions}

As discussed above, the distribution of parameter space points for which dangerous short-lived false vacua
occur in the 2HDM is not uniform. The regions of parameter space which conform to equations~\eqref{eq:M0},
\eqref{eq:astr} or have the discriminant~\eqref{eq:disc} negative are usually not easily visualized in 2-dimensional
slices, and as explained in section~\ref{sec:mindeg}, the tunneling time to lower vacua may depend
crucially on the landscape of saddle points in field space -- and this will also depend in a non-trivial
manner on the numerical values of the couplings, affecting the number of possible solutions of the minimisation
conditions of eq.~\eqref{eq:min}. In the present subsection, we will give two examples of parameter choices
for which, due to different reasons, the EW vacuum is perfectly safe. In all cases to follow we study
model type I, and
fix six of the parameters of the model, allowing two others to vary in such a manner as to comply
with theory and experimental constraints. Since we wish to have a physically interesting EW vacuum, we
chose to specify the values of (in principle) observable 2HDM parameters, rather than the couplings of the
potential in eq.~\eqref{eq:pot2}. To that end, we of course chose the value of $v = 246$ GeV and
$m_h = 125$ GeV for the EW vacuum, and then proceed to select, for each benchmark scenario, the masses
$m_H$ and $m_{H^\pm}$, the value of $\tan\beta$ and of $\sin(\beta-\alpha)$ -- thus, indirectly, the value of
$\alpha$. We chose $\sin(\beta-\alpha)$ because this quantity is already quite constrained to be close to unity
by LHC data.

The 2HDM parameter scan we undertake therefore considers these six parameters fixed and then proceeds
to choosing random values for two others, which we chose to be $\lambda_5$ and $m_A$~\footnote{The
reason for this is related to eqs.~\eqref{eq:coup}, which show it to be an efficient choice of
parameters to fully specify the 2HDM potential.}. Each selection of parameters is then verified for
theory and experimental constraints, and if all are obeyed a satisfactory EW vacuum is found.
{\em A posteriori} the existence of a second minimum is checked, and if that second minimum is the global
one, the tunneling time to the true vacuum is computed.

\begin{itemize}
\item Decoupling scenario
\end{itemize}

As a first example, we have chosen $m_H = 600$ GeV, $m_{H^\pm} = 700$ GeV, $\tan\beta = 1$ and
$\sin(\beta-\alpha) = 0.99$. Though $\lambda_5$ was allowed to vary between -10 and 10, only values in
the window between $\sim$ -6.3 and $\sim$ -2.3 were found after all constraints applied~\footnote{Notice how the
exclusion bounds presented in fig. 6 of~\cite{Aaboud:2018eoy} come from a 2HDM analysis which chooses values for the
soft breaking such that $\lambda_5 = 0$. Though direct application of those exclusion bounds to our
results is not possible, the model independent cross section limits presented in the afore-mentioned reference
may exclude some of the parameter space discussed here.}. Likewise,
the pseudoscalar mass is found to be constrained between roughly 620 and 705 GeV. It is well known
that the electroweak precision constraints (namely the bounds on the Peskin-Takeushi parameters
$S$, $T$ and $U$) force the extra scalar masses to be close in the high mass range, so these
results are not surprising.

The high values for the extra scalar masses coupled with the fact that $\sin(\beta-\alpha)$ is extremely
close to 1~\footnote{This implies that the coupling of $h$ to Z or W bosons and to fermions is very much SM-like.},
means we are well within a decoupling regime for the model. And of course, one of the possible explanations
for the current LHC results is the decoupling of all BSM particles, yielding a SM-like 125 GeV scalar. Thus
the benchmark scenario chosen herein is of experimental interest.

What then can we conclude regarding the stability of the EW vacuum for this benchmark scenario? For all
200000 points generated complying with the choices above for the parameters and all constraints, we
observe that the conditions for the possible existence of two minima, eqs.~\eqref{eq:M0} and~\eqref{eq:astr},
are {\em never} satisfied. Thus, for this benchmark scenario, the EW vacuum is unique, and thus (at tree-level
at least) entirely stable. Please beware: this does not mean that any choice of parameters in the decoupling
regime will always fall into this category, although, as explained above for the type II model, large
masses tend to yield stable EW vacua. Of course, there is no need to go into the decoupling regime
to find parameters for which no non-degenerate vacua do not exist, the blue points in fig.~\ref{fig:types}
show this to be true.
\begin{figure}[t!]
	\centering
	\includegraphics[width=0.6\linewidth]{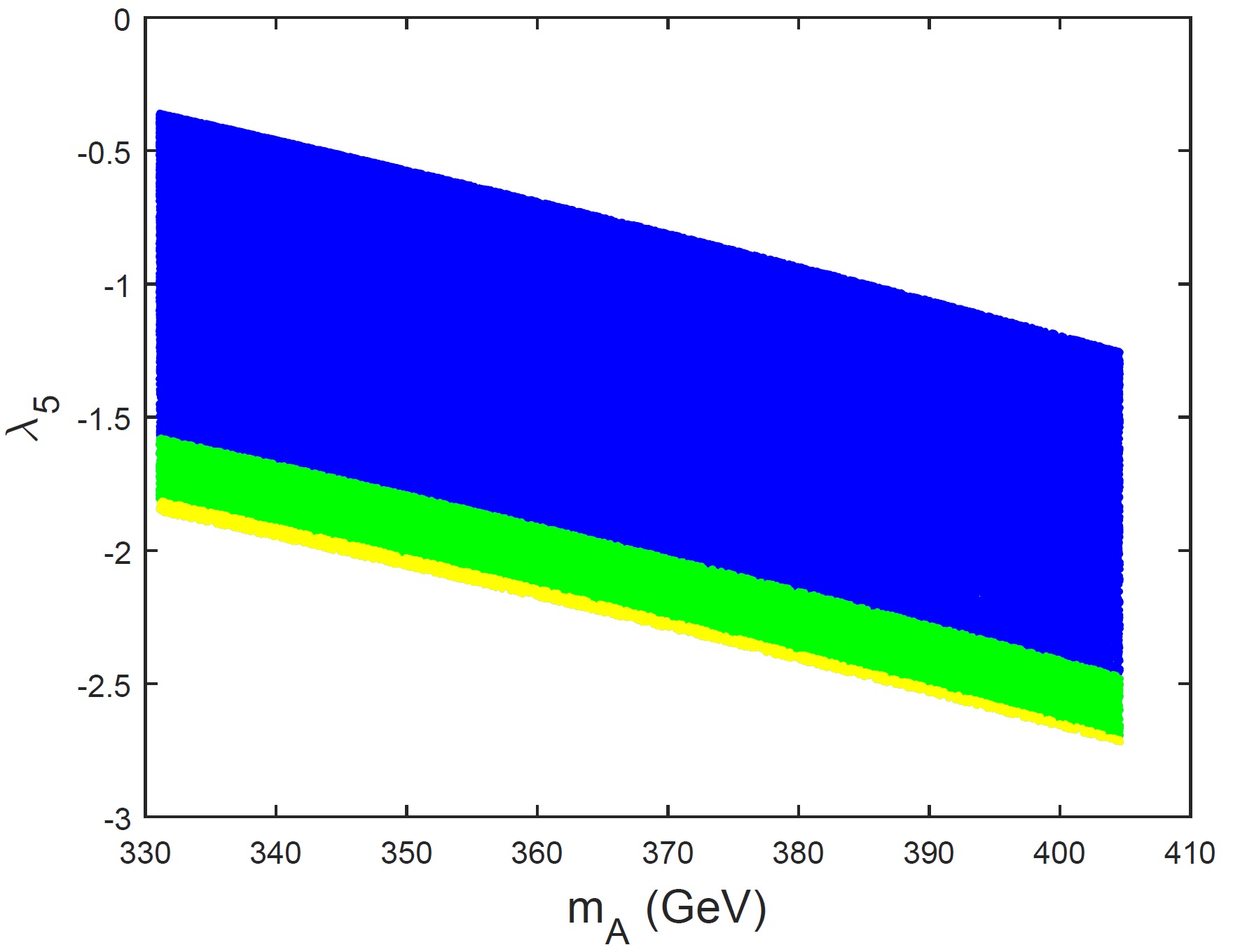}
	\caption{Scatter plots of $\lambda_5$  {\em vs.} $m_A$
		for the low mass stable benchmark scenario considered. The colour code is the same as in fig.~\ref{fig:types}.}
	\label{fig:bm1}
\end{figure}
Thus some regions of 2HDM parameter space have EW vacua which are unique at tree-level. Therefore, apart from
the possibility of one-loop corrections to the potential originating deeper vacua as seems to be the case in the SM,
the stability of the EW vacuum  in such 2HDM parameter space regions is ensured and no tunneling calculations
are needed.

\begin{itemize}
\item Low mass stable scenario
\end{itemize}

Consider now a different choice of parameters corresponding to much lower masses for the extra scalars:
$m_H = 280$ GeV, $m_{H^\pm} = 400$ GeV, $\tan\beta = 2.3$ and again $\sin(\beta-\alpha) = 0.99$. This last
choice all but ensures $h$ has SM-like behaviour. The value of $\tan\beta$ is chosen such as to comply with
the exclusion in the $\tan\beta$--$m_{H^\pm}$ plane stemming from B-physics constraints~\cite{Arbey:2017gmh}.
And the low masses chosen for $H$ and $H^\pm$ are obviously interesting from the experimental point of view,
as they raise the possibility of new particles discovered at LHC. As before, electroweak precision
constraints force the pseudoscalar to be close in mass to the charged scalar, as can be appreciated from
fig.~\ref{fig:bm1}. In this plot we show a ``phase diagram" of the 2HDM parameter space. Unlike the plots
in fig.~\ref{fig:types} -- the colour code is the same here than in those plots -- the parameter
space now being scanned is truly a two-dimensional one, and thus fig.~\ref{fig:bm1} gives us a clearer
picture of regions having different vacuum structure.

What we observe in fig.~\ref{fig:bm1} is the total absence of red points, and only a thin yellow strip where
the EW vacuum could be the false vacuum. Indeed, for all points for which the EW vacuum is indeed a local
minimum of the potential and not the global one, tunneling time calculations have revealed that the lifetime of the false
vacuum is always far superior to the current age of the universe. Thus, even though for this benchmark scenario
there may be dual minima, and ``our" vacuum is not guaranteed to be the true vacuum of the model, it is nonetheless
found to be either stable or incredibly long lived. Hence one must be careful to {\em not} exclude offhand
regions of parameter space for which the discriminant $D$ from eq.~\eqref{eq:disc} is negative -- $D>0$ is a
necessary and sufficient condition for the EW vacuum with $v = 246$ GeV to be the global minimum of the theory, but
as this example shows, points with $D<0$ may be entirely acceptable, having lifetimes larger than $T_U$. One must
therefore be cautious in excluding regions of parameter space using the sign of discriminant $D$, as was made in
refs.~\cite{Staub:2017ktc,Basler:2017nzu}: if $D<0$, tunneling times involving two-field bounce equations need to be computed,
lest one is needlessly refusing phenomenologically acceptable combinations of 2HDM parameters.

\subsection{Second benchmark scenario: considerable exclusion}
\label{sec:decad}

\begin{figure}[ht]
  \centering
  \includegraphics[width=0.47\linewidth]{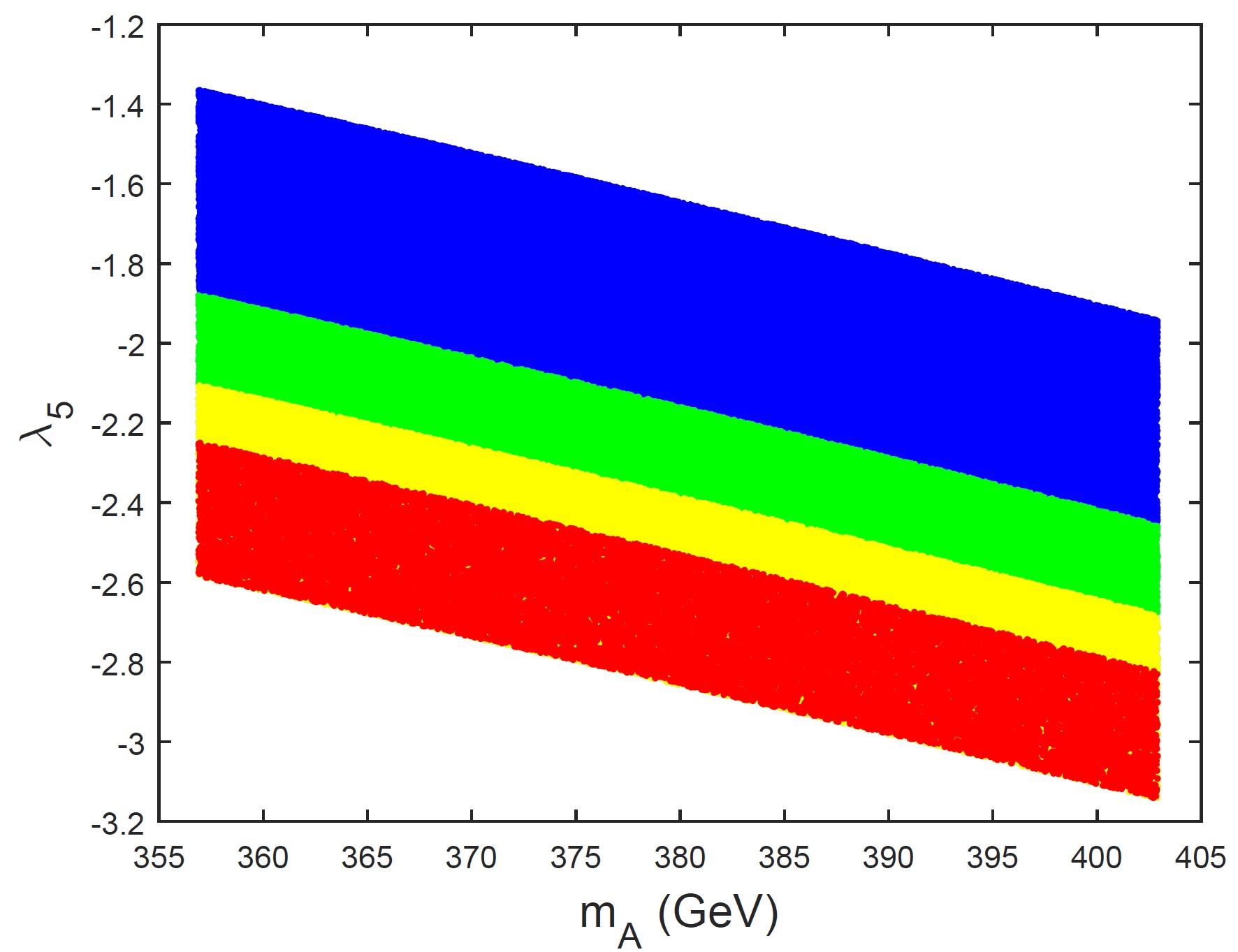}
  \includegraphics[width=0.47\linewidth]{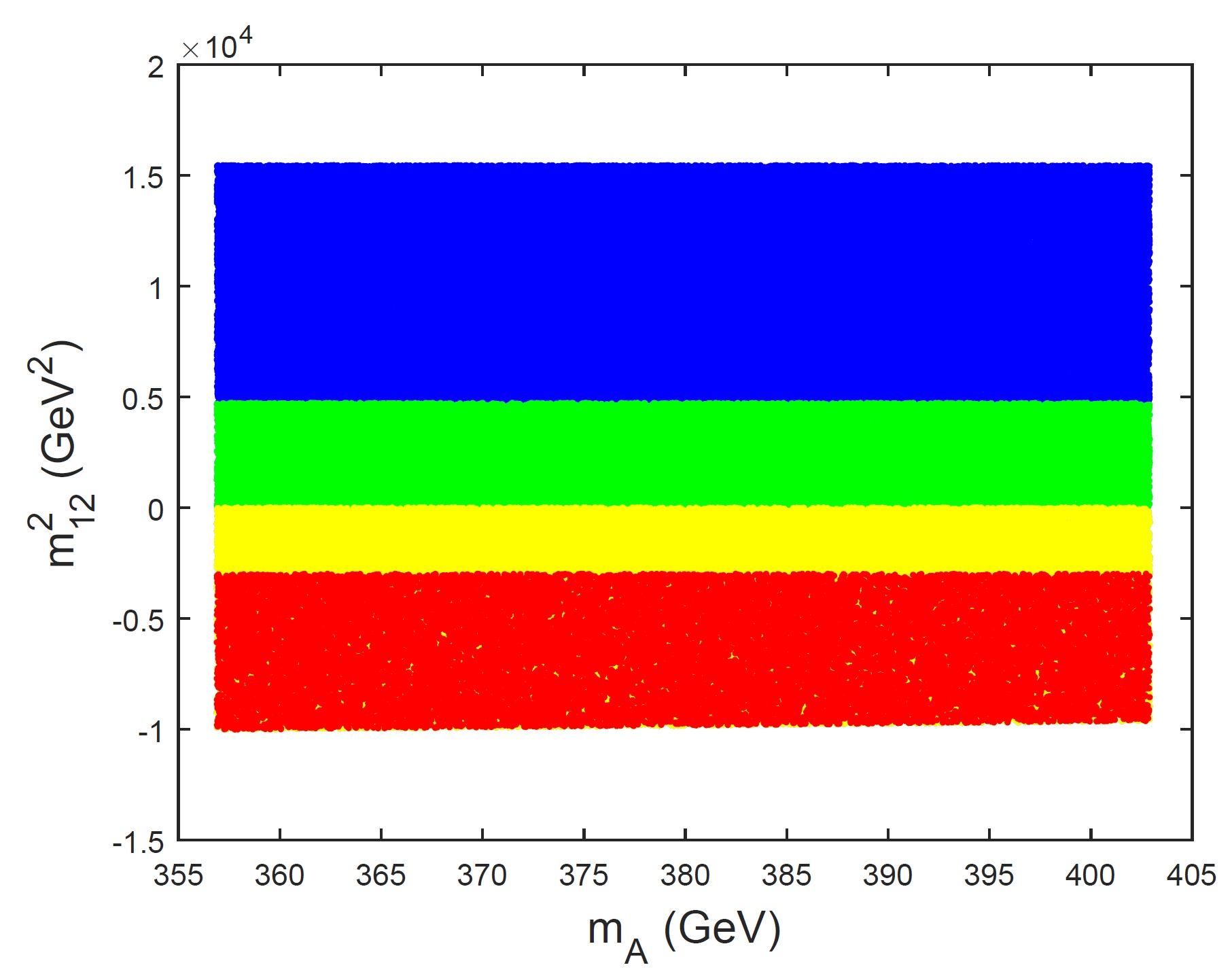}
  \caption{Scatter plots of $\lambda_5$ (left) and $m^2_{12}$ (right) {\em vs.} $m_A$
  for the benchmark scenario considered in section~\ref{sec:decad}. The colour code is the same as in fig.~\ref{fig:types}.}
 \label{fig:bm2}
\end{figure}

The vacuum stability of the 2HDM may however change dramatically even for seemingly small variations
in its parameters. Consider yet another choice of parameters, still corresponding to low masses for the extra scalars:
$m_H = 200$ GeV, $m_{H^\pm} = 400$ GeV, $\tan\beta = 2.5$ and again $\sin(\beta-\alpha) = 0.99$. Though this
choice of parameters seems to be very similar to the previous benchmark considered, the allowed vacuum structure of the
model is now quite different, as may be appreciated from fig.~\ref{fig:bm2}. Notice how the region where two minima
are in principle allowed (the yellow points) is now much larger, and how many red points now occur. In fact, for
this region of parameter space, roughly 67\% of all cases where ``our" vacuum is the higher minimum yield
a tunneling time inferior to the age of the universe. Globally, we find that for the points generated in this
benchmark scenario which verify all theory and experimental constraints, roughly 11\% have dangerous tunneling times.
This is a much greater percentage than the one found for the blind parameter scan, showing that specific regions
of the parameter space may be much more prone to dangerous false vacua than others. It is quite stunning
how merely increasing by 0.2 the value of $\tan\beta$ and reducing by 80 GeV the value of the heavier CP-even
scalar may have such a drastic effect in the vacuum structure of the 2HDM, but that simply reflects the
complicated and non-obvious dependence on these parameters in eqs.~\eqref{eq:min},~\eqref{eq:M0},~\eqref{eq:astr}
and~\eqref{eq:disc}. This not to mention the susceptibility of the tunneling time calculations to the geometry of
the potential -- which may be heavily influenced by changes in the potential's couplings -- as illustrated in
figs.~\ref{fig:lands} and~\ref{fig:val}. The reader should also consider the importance of calculating the tunneling rates
to the degenerate true vacua -- as discussed in section~\ref{sec:mindeg}, despite that degeneracy originating physically
equivalent vacua, the lifetime of the false vacuum can change immensely if one does not take into account the
existence of two possible true vacua it can decay into. In this present case, doing the lifetime calculation
correctly taking into account both lower vacua yielded roughly 50\% more dangerous red points than if we considered
tunneling to only one of the lower vacua.

In the right of fig.~\ref{fig:bm2}
we plot $m^2_{12}$ against $m_A$, illustrating nicely how all EW false vacua can only occur for negative
values of the soft breaking parameter. This is a known feature of the 2HDM -- negative discriminant $D$ seems to
only occur for negative $m^2_{12}$, though there is no demonstration of this property. It would imply a correlation
in the signs of the two last terms in the definition of the discriminant in eq.~\eqref{eq:disc}.

\section{Conclusions}

The 2HDM has a rich vacuum structure, with the possibility of coexisting electroweak breaking, CP conserving,
non-degenerate vacua already at tree-level. Instability of a phenomenologically desirable vacuum, where both doublets
have vevs $v_1$ and $v_2$ such that $v_1^2 + v_2^2 = (246$ GeV$)^2$ and elementary particles have their known masses,
is therefore a possibility in the 2HDM. The SM vacuum instabilities discussed in the literature are of a different
nature -- they occur due to radiative corrections to the potential, whereas in the 2HDM they may already arise
at tree-level. Another difference in the 2HDM case involves the larger number of scalar fields which the potential
possesses, which complicate considerably the calculation of the lifetime of eventual false vacua -- clearly, a local
minimum with a lifetime superior to the age of the universe should not, in principle, be excluded from consideration,
since it might yield an acceptable description of known phenomenology.

In this work we analysed in depth the possibility of using the lifetime of false vacua as an exclusion tool of
regions of 2HDM parameter space. The gauge freedom of the model allowed us to reduce the complexity of an {\em a priori} 8-field
problem -- and the physics of the models under discussion, coupled with the shape of the
bounce equations describing the tunneling trajectories between vacua, permitted a further simplification.
Vacua of different kinds cannot coexist in the 2HDM -- existing theoretical results have proved that, for
CP-conserving models, only two non-degenerate CP-conserving minima, both containing  real vevs, can coexist.
In models with explicit CP violation, on the other hand, the CP symmetry is not defined from the start, and minima
with and without complex vevs can also coexist. We have shown that, in the CP-conserving 2HDM potential, the bounce
equations, coupled with the necessary boundary conditions for the bounce, and the structure of the potential's derivatives
dictated by CP symmetry, reduce the tunneling time calculation
to a 2-field problem. The remaining fields do not contribute, their
bounce equations only allowing for trivial, vanishing solutions when the appropriate boundary conditions are taken
into account. There are two crucial points in this reduction to a 2-field problem: (a) the bounce equations factorize
on the fields whose vevs vanish at either minima; (b) at both the true and false vacua, only two fields have
non-vanishing vevs. Both of these points are direct consequences of the physics involved, in this
case the CP-conserving structure
of the potential. On the other hand, for an explicitly CP-violating 2HDM potential, the factorization in the imaginary
component field of one of the doublets no longer occurs, and minima with real and complex vevs can now coexist. Bounce
solutions involving three fields become possible, something which is yet again a direct consequence of the physics
involved, not simply a mathematical property.

Gauge freedom, the CP symmetry, the bounce equations and boundary conditions imposed by physics-limited vacuum
structure of the 2HDM therefore allow us to reduce the computation of the tunneling time to a 2-field
problem. However, the 2-field case produces a bizarre consequence. The 2HDM is invariant under a simultaneous
sign redefinition for both scalar doublets, and no physical consequences should in principle arise from such a sign
swap. Indeed, for any pair of solutions of the minimisation conditions of the potential,
its symmetric is also a solution. This is a well-known, and trivial, property of 2HDM vacuum
solutions. Any minimum found is in fact a ``pair of minima", degenerate, separate in field space, each producing
exactly the same physics. Hitherto the existence of such degenerate pairs of minima has been seen as nothing more
than a curiosity, but here we have shown it may have significant impact in the lifetime of false vacua: indeed,
the false vacuum can decay to a pair of degenerate true vacua, separated in field space, and the trajectory to
each of the true vacua will not be, in general, the same. Hence the partial decay rates to each of the deeper
vacua will in general be different, and the landscape of maxima and saddle points found along the trajectories
to each true vacuum can indeed yield vastly different tunneling rates. We have found many instances where considering
both tunneling possibilities yielded false vacua with lifetimes shorter than the age of the universe -- whereas
considering only one of the decaying possibilities seemed to indicate a stable false vacuum.

Applying the theoretical insights gained on generic scans of the 2HDM parameter space, we analysed which regions
of that parameter space might be excluded on grounds of short EW vacuum lifetime. Generic
scans over all allowed -- under theory and experimental constraints -- parameters show that the existence of
non-degenerate minima is rare in the 2HDM, and that even when a false vacuum occurs, its lifetime is
often found to be superior to the age of the universe. The percentage of 2HDM parameter space points
excluded in generic scans is then found to be of the order of a few percent. However, care must be exercised in
reading this result, since the regions of 2HDM parameter space where non-degenerate minima occur are not
uniformly distributed -- and neither is the subset of those for which short-lived false vacua occur. We therefore
proceeded to considering specific benchmark scenarios, illustrating how three very different regimes might occur.
First we considered a choice for extra scalar masses and angles $\alpha$ and $\beta$ that pushed the theory well
into the decoupling regime. Such a choice corresponds to a region of parameter space for which no non-degenerate
minima exist in the potential, and as such the model is entirely stable at tree-level. The decoupling regime, of
course, is not the only case where no non-degenerate vacua do not occur, the same does happen for smaller masses
of the extra scalars. The second scenario considered studied a low-mass case for the extra scalars, and for which
the possibility of a false EW vacuum now arises -- certain regions of the considered parameter space had
$D < 0$, the discriminant which characterizes false vacua. However, for all such false vacua, the tunneling times
towards the true vacuum were always found to be larger than $T_U$, and therefore stability
is ensured. Thus the mere existence of a false vacuum should not be used {\em per se} to exclude regions of parameter
space for which $D < 0$ -- tunneling times should and must be computed, and parameter exclusion should only be decided
after that calculation.

Finally, we considered a low-mass scenario for which a large swath of parameter space is excluded on grounds of
the short lifetime of the false vacuum found. The importance of a proper lifetime calculation -- taking into
account the existence of a pair of lower true vacua, related by sign changes in the values of the vevs -- was
emphasised. In fact, the number of dangerous vacua found can increase by as much as 50\% when the full
vacuum structure is taken into account. Though dangerous vacua are hard to pinpoint in terms of relations
between potential couplings or physical observables, we observed that a negative discriminant only seems to
occur for a negative soft breaking term $m^2_{12}$.

The overall conclusion of this work is that 2HDM vacuum instability at tree-level can have significant impact
on parameter exclusion for certain regions of the parameter space -- but that requires an appropriate calculation
of the bounce solutions, taking into account the 2-field dynamics that CP-conservation allows us to study. Of crucial
importance is also the seemingly trivial existence of pairs of degenerate, sign-swap-related, true vacua, since
the lifetime of the false vacua may depend enormously on that fact. Generic blind scans of 2HDM parameters may
suggest that the frequency of dangerous vacua is very small, but we have shown that they may be quite abundant for
specific, experimentally-interesting, regions of parameter space. Though the current analysis was performed at
tree-level, the significance of the results found is undeniable. Of course, from the existing SM studies, we can
expect that radiative corrections will further complicate matters and bring more possibilities of vacuum instability.
Thus a one-loop extension of the present work should be undertaken. It should also be considered, though, that unlike
the SM, for which the top quark induces loop instabilities, in the 2HDM the larger scalar content of the model
does in some cases counteract the fermion sector instability -- it has been shown that the 2HDM, with a single
tree-level vacuum, can be stable all the way up to the Planck scale, unlike what is claimed for the
SM (see, for instance, \cite{Ferreira:2015rha,Basler:2017nzu}). In which case,
though higher order corrections to the present work are welcome, it may be that they do not bring major differences
to the results shown here.

\section*{Acknowledgments}
The work of VB and FC is carried out within the INFN project QFT-HEP. PF would like to thank the gratious hospitality and financial support from INFN, Sezione Catania, where this work was partially developed, as well as partial support under CERN fund grant CERN/FIS-PAR/0002/2017. This work is also partially supported by the Polish National Science Centre HARMONIA grant under contract UMO-2015/18/M/ST2/00518 (2016-2019).

\vspace*{1cm}
\bibliography{tunnel}
\end{document}